\documentclass[]{aa}

\usepackage{tikz}
\usepackage{xcolor}

\usepackage{pgf}	

\usepackage{txfonts}
\usepackage{natbib}
\usepackage{epsf}
\usepackage{rotating}
\usepackage{graphics}
\usepackage{hyperref}
\usepackage{verbatim}
\usepackage{longtable}
\usepackage{multirow}
\usepackage{graphicx}
\usepackage{booktabs}
\usepackage[font=small,labelfont=bf]{caption}

\def \ms{m\,s$^{-1}$}
\def \1s{$1\,\sigma$}

\def \t0{T$_0$}

\DeclareUnicodeCharacter{2212}{ --}

\begin{document}

\title{Weighing the mass of LHS 3844 b}

\author{
A.~Hacker\inst{1} \and
N.~Astudillo-Defru\inst{2}\and
R.~F.~D\'iaz\inst{1,3} \and
C.~Dorn\inst{4}\and
X.~Bonfils\inst{5}\and 
J.M.~Almenara\inst{5,6}\and 
P.~Cortés-Zuleta\inst{7}\and
X.~Delfosse\inst{5}\and 
T.~Forveille\inst{5}\and 
S.~Udry\inst{6}
}

 \offprints{A. Hacker (ahacker@unsam.edu.ar), N.~Astudillo-Defru (nastudillo@ucsc.cl)}

\institute{
Instituto de Ciencias F\'isicas (CONICET / ECyT-UNSAM), Campus Miguelete, 25 de Mayo y Francia, (1650) Buenos Aires, Argentina.\and
Departamento de Matem\'atica y F\'isica Aplicadas, Universidad Cat\'olica de la Sant\'isima Concepci\'on, Alonso de Rivera 2850, Concepci\'on, Chile. \and
Instituto Tecnol\'ogico de Buenos Aires (ITBA), Iguaz\'u 341, Buenos Aires, CABA C1437, Argentina. \and
Institute for Particle Physics and Astrophysics, ETH Zürich, 8093 Zürich, Switzerland.\and
Univ. Grenoble Alpes, CNRS, IPAG, 38000 Grenoble, France.\and
Observatoire de Gen\`eve, Département d’Astronomie, Universit\'e de Gen\`eve, Chemin Pegasi 51b, 1290 Versoix, Switzerland.\and
SUPA, School of Physics and Astronomy, University of St Andrews, North Haugh, St Andrews KY16 9SS, UK.
}

\date{Received TBC; accepted TBC}
      
\abstract
{LHS~3844~b (TOI-136\,b) is a ultra short-period, Earth-size exoplanet detected by TESS. It is one of the most favourable object for atmospheric characterisation and the study of its surface with the James Webb Space Telescope. 
However, the dynamical mass of this planet has not been measured yet.}
{We aim to determine the mass of LHS~3844~b using high-precision radial velocity (RV) measurements and assess the robustness of the inferred signal across different noise and orbital modelling assumptions.}
{We analyse 25 ESPRESSO RV observations within a fully Bayesian framework. We explore 15 competing RV models that differ in their treatment of correlated stellar variability (through different Gaussian Process kernels) and long-term drifts. Marginal likelihoods are computed for all models and used for Bayesian model comparison and evidence-weighted parameter estimation.}
{The RV planetary signal is robustly detected across all models, and the inferred semi-amplitude remains stable under all tested noise and drift prescriptions. From the evidence-weighted posterior samples we derive a planetary mass of $2.27 \pm 0.23$ M$_\oplus$ and a bulk density of $5.67 \pm 0.65$ gcm$^{-3}$, consistent with a predominantly rocky composition. Model comparison favours GP kernels including periodic or quasi-periodic components associated with stellar rotation and disfavours models with additional long-term drifts. Using interior-structure inference, we find that the core mass fraction is comparable to (or slightly smaller than) Earth’s and only trace amounts of water are permitted, supporting a dry, terrestrial interior. We also investigate a tentative additional signal near $\sim6.9$ days, but Bayesian model comparison does not provide conclusive support for its planetary interpretation.}
{We report the first dynamical mass measurement of LHS~3844~b, confirming it as a dense, terrestrial ultra–short-period planet. With its exceptionally well-constrained bulk properties and extensive JWST programme, LHS~3844~b remains a benchmark target for studies of atmospheres and surfaces of rocky exoplanets.}

\authorrunning{Hacker et al.}
\titlerunning{Weighing the mass of LHS 3844 b}

\maketitle

\section{Introduction \label{sect.intro}}
Detecting and characterizing Earth-like planets is a key step toward understanding planetary formation and evolution. Photometric surveys, such as the Transiting Exoplanet Survey Satellite (TESS) mission \citep{ricker2014}, have identified hundreds of Earth-sized planets and candidates, many of which are subsequently characterised with ground-based radial velocities measurements \citep[e.g.][]{hacker2024,serrano2024,armstrong2025}. Ultra-short-period planets (USPs) provide a unique opportunity for precise mass and radius measurements, as their RV semi-amplitudes are typically an order of magnitude larger than those of longer-period terrestrial planets \citep{sanchis-ojeda2014}. Additionally, USPs allow for the development of specific observational techniques to efficiently mitigate the effects of stellar rotational modulation \citep{hatzes2011}. Accurately determining their fundamental parameters enables the measurement of bulk densities, which is crucial for constraining planetary compositions and internal structure. Understanding the compositions of these planets, in turn, informs models of planetary formation and evolution.

\par USPs are terrestrial worlds with orbital periods shorter than one day \citep{sahu2006,sanchis-ojeda2014,winn2018}. They are relatively rare, occurring around $\sim$1\% of stars \citep{sanchis-ojeda2014,petrovich2019}, and are generally smaller than 2 $R_\oplus$ \citep{rappaport2013}. USPs are typically found in multi-planet systems, favouring hosts with solar or subsolar metallicity \citep{winn2017}. 
Their architectures often show enhanced mutual inclinations compared to compact systems of longer-periods, suggesting dynamical excitation accompanying orbital shrinkage \citep{dai2018}. USPs are also believed to have undergone significant migration, as their present-day orbits lie within the dust sublimation radius for solar-type hosts \citep{millholland2020}, and recent statistical studies indicate that their occurrence increases with stellar age and that the architecture of USP systems evolves on gigayear timescales, consistent with inward tidal migration \citep{tu2025}. Their bulk densities suggest compositions consistent with terrestrial planets, often resembling Earth-like or iron-rich interiors \citep{rappaport2013,dai2019,uzsoy2021}. Given their extreme irradiation (>650 $S_\oplus$ for Sun-like hosts), USPs are believed to be stripped rocky cores \citep{sanchis-ojeda2014,lundkvist2016}, a hypothesis supported by transmission and emission spectroscopy \citep{crossfield2022,kreidberg2019,zhang2024}. Precise measurements of mass and radius are particularly important in this context, as they allow for direct tests of atmospheric loss mechanisms and internal differentiation processes.

\par While most USPs lack significant volatile envelopes, some larger ones (e.g., TOI-561 b and 55 Cnc e) exhibit anomalously low densities that suggest the possible presence of residual atmospheres \citep{brinkman2023,demory2016,tsiaras2016,dai2019}, and recent observations indeed indicate a secondary atmosphere in 55~Cnc~e \citep{hu2024}.
Precise composition constraints on these planets provide valuable insight into the mass dependence of planetary differentiation, particularly during the final stages of the giant impact phase \citep{scora2022}. Furthermore, their short orbital periods make them ideal candidates for detailed surface characterisation with the James Webb Space Telescope (JWST), through the analysis of the phase light curve. 

One of the earliest planetary discoveries from TESS was LHS 3844b, a highly irradiated terrestrial exoplanet orbiting the nearby mid-M dwarf LHS 3844 (\citealt{vanderspek2019}). Located 15 parsecs away, this USP planet completes an orbit in approximately 11 hours and has a radius of $1.303 \pm 0.022 R_{\oplus}$, placing it within the terrestrial planet regime. LHS 3844b is a key target for atmospheric and compositional studies because it orbits a small ($0.189 \pm 0.006\,R_{\odot}$), nearby star, making it a spectroscopically accessible terrestrial exoplanet.

The significance of LHS 3844b as a target for exoplanetary studies was significantly raised by multiple observational and theoretical efforts. Infrared phase curve measurements from \textit{Spitzer} indicated that the planet does not have a thick atmosphere, down to a limit of 10 bar, consistent with efficient atmospheric erosion over its lifetime \citep{kreidberg2019}. This conclusion was reinforced by \citet{diamondlowe2020}, who conducted ground-based optical transmission spectroscopy with Magellan/LDSS3C and found a featureless spectrum, disfavouring clear, low mean molecular weight atmospheres at surface pressures of 0.1 bars or greater. The role of high-energy stellar irradiation in driving atmospheric loss has been investigated, with estimates suggesting that the high XUV irradiation from LHS 3844 would have stripped a primordial hydrogen-helium atmosphere within 54 Myr \citep{diamondlowe2021}. In contrast, \citet{kane2020} explored an alternative scenario based on stellar age estimates ($7.8 \pm 1.6$ Gyr) and geodynamical models, concluding that the volatile-poor nature of LHS 3844b may be the result of in situ formation inside the system’s snow line rather than post-formation atmospheric loss, although a giant impact stripping the planet's atmosphere is also possible.
More recently, \citet{kokori2023} refined the transit ephemeris of LHS~3844b, providing updated estimates of the orbital period and mid-transit time, $P_{\mathrm{t}} = 0.46292964 \pm 0.00000033$ d and $t_{0\mathrm{t}} = 2458828.93037 \pm 0.00025$ d (BJD), which we adopt in this work.

\par The planet has also been a prime target of the JWST. Over 40 hours were granted during Cycle 1 and 2 to study the planet mid-infrared spectrum with MIRI\footnote{\url{https://www.stsci.edu/jwst/science-execution/program-information?id=1846}} and the planetary surface roughness with NIRSpec.\footnote{\url{https://www.stsci.edu/jwst/phase2-public/4008.pdf}} 

Despite the extensive observational efforts dedicated to LHS 3844b, a key piece of information remains unknown: its mass. A precise mass measurement is essential to constrain the planet’s bulk composition and to determine whether it aligns with the terrestrial planets of the Solar System or represents a different class of rocky world. Moreover, an independent determination of the surface gravity is critical for interpreting atmospheric data gathered in transmission, as it directly influences atmospheric scale height and composition retrievals \citep[e.g.][]{heng2017}. Measuring the dynamical mass of LHS 3844b would therefore provide a fundamental constraint for both interior and atmospheric studies, refining our understanding of terrestrial planets around M dwarfs. In this work, we present the first mass determination of LHS 3844b, placing it in context with the growing population of well-characterised terrestrial exoplanets. 

In Sect.~\ref{sect.data} we describe the observational data and its reduction process.  Sect.~\ref{sect.stellar} deals with the characterisation of the central star of the system. In Sect.~\ref{sect:preliminary} we perform a preliminary exploration of the radial velocity data, including a GLS periodogram analysis and tests of additional linear covariates and mild orbital eccentricity. 
Sect.~\ref{sect.rvanalysis} describes our radial velocity modelling strategy, consisting of a Bayesian analysis of 15 alternative models combining different Gaussian Process covariance structures and prescriptions for long-term velocity drifts.
The results of the model comparison and the inferred planetary and orbital parameters are presented in Sect.~\ref{sect.discussion}, including an assessment of additional periodic signals and an interior characterisation of the planet. 
Finally, Sect.~\ref{sect.conclusion} summarises our main findings and conclusions.

\section{Observations and data \label{sect.data}}
\subsection{Observations}
LHS3844 was observed with the Échelle SPectrograph for Rocky Exoplanets and Stable Spectroscopic Observations (ESPRESSO), a fibre-fed \emph{echelle} spectrograph installed at the European Southern Obvservatory Very Large Telecospe (VLT) of the Paranal Observatory. A technical description of the instrument is provided by \citet[][and references therein]{megevand2014, pepe2014b} and its on-sky performance is evaluated by \citet{pepe2021}. The spectrograph works between 378.2 nm and 788.7 nm, and has a resolving power between 130,000 and above 190,000, depending on the observing mode, when the instrument is used in single-telescope mode\footnote{ESPRESSO possesses a operational mode that can combine light from all four Unit Telescopes at the VLT.}. \citet{pepe2021} demonstrated that ESPRESSO provides radial velocity measurements precise at the level of 10 cm s$^{-1}$. Since its commissioning, ESPRESSO has been used in a large number of studies \citep[e.g.][]{faria2020, hobson2024, suarezmascareno2020, suarezmascareno2024, ramirez2025, cristiani2025}

LHS 3844 was targeted over two seasons: from November 2nd 2018 to December 3rd 2018 (14 visits; program 0102.C-0496, PI:Astudillo-Defr\'u) and from June 9th 2019 to September 6th 2019 (11 visits; program 0103.C-0849,  PI:Astudillo-Defru). In total 25 spectra were acquired, with a exposure time of 2400 seconds for each in slow readout mode, and using 2x1 binning (mode HR21) of the instrument. In this mode, the spectrograph resolving power is around 138,000. The second fibre was used for sky subtraction. 

In June 2019, the ESPRESSO fibre link was upgraded, which likely introduced a radial velocity offset. We therefore treat data before and after the upgrade separately, allowing for a velocity offset between them, as suggested by \citep{pepe2021}. There are 15 and 10 measurements taken before and after the upgrade, respectively. The 15 ESPRESSO-pre-upgrade spectra reached at 610 nm a median signal-to-noise ratio (SNR) of 16.1, with a minimum SNR of 11.8 and a maximum SNR of 18.7. The 10 ESPRESSO-post-upgrade spectra reached a median signal-to-noise ratio (SNR) of 19.6, with a minimum SNR of 15.4 and a maximum SNR of 24.3 at the same wavelength.

\subsection{Data reduction and exploratory analysis}

The ESPRESSO spectra were reduced using the ESO data pipeline version 3.0.0, described in \citet{pepe2021}. The extracted wavelength-calibrated two-dimensional spectra resulting from the extraction were used to measure the radial velocities using a template matching technique \citep{astudillodefru2015, astudillodefru2017b}. Due to the ESPRESSO intervention, we processed data from pre- and post-intervention as independent data set. The S/N ratio reached translates into a median photon noise uncertainty of 60 cm s$^{-1}$ and 52 cm s$^{-1}$, for the velocities before and after the upgrade operation, respectively, in line with the performances of ESPRESSO described in \ref{sect.data}.

In addition to the radial velocities measured with the template matching technique, we computed the cross-correlation function (CCF) of the spectra with a numerical mask \citep{baranne1996, pepe2002}, and measured the CCF full-width at half-maximum (FWHM) and contrast. We also measured activity indexes from the ESPRESSO spectra. The H$\alpha$ index was computed following the prescriptions of \citet{cincunegui2007} and \citet{DaSilva2011}. The H$_\beta$, H$\gamma$ and Na~D indices were computed using fluxes integrated within narrow wavelength bands centered on the respective spectral features, which are summarized in Table~\ref{tab:activity_indices}.

\begin{table}[ht]
\centering
\caption{Definition of the wavelength bands used for the computation of the activity indices. 
}
\label{tab:activity_indices}
\begin{tabular}{lcc}
\hline
\textbf{Index} & \textbf{Band} & \textbf{Wavelength range [\AA]} \\
\hline
H$_\beta$ & Central (C) & 4861.04--4861.60 \\
          & Violet (V)  & 4855.04--4860.04 \\
          & Red (R)     & 4862.60--4867.20 \\
\hline
H$\gamma$ & Central (C) & 4340.162--4340.762 \\
          & Violet (V)  & 4333.60--4336.80 \\
          & Red (R)     & 4342.00--4344.00 \\
\hline
Na~D      & D$1$ & 5895.92 $\pm$ 0.25 \\
          & D$2$ & 5889.95 $\pm$ 0.25 \\  
          & Violet (V)  & 5860.0--5870.0 \\
          & Red (R)     & 5904.0--5908.0 \\
\hline
\end{tabular}
{\captionsetup{font=small}
\caption*{\textbf{Notes.} For H$_\beta$ and H$\gamma$, the index is defined as $C / (V + R)$, and for Na~D as $(\text{D1} + \text{D2}) / (V + R)$.}}

\end{table}

The resulting dataset is presented in Table~\ref{tab:data_summary}. None of the observations is expected to be contaminated by moonlight. In all observations, the Moon was farther than 50 degrees from the target star. Also, no measurements are expected to be contaminated by sunlight either, because no observation was made during twilight. However, an exploratory data analysis revealed an unusually small FWHM for the observation on night BJD 2\,458\,643.91, the last visit before the fibre upgrade. For this observation, the contrast was not computed by the pipeline. This observation was performed closest to the morning twilight, but the Sun was still more than 21 degrees below the horizon. No contamination is expected, and no trend in velocities is seen with Sun altitude, but the coincidence is noteworthy. A deeper study of the effect of sunlight on ESPRESSO observations is beyond the scope of this paper.
The secular acceleration was computed from LHS 3844 proper motion and parallax \citep{gaia3edr}, yielding 0.219 m/s/yr, equivalent to 0.185 m/s over the time span of observations. The effect of secular acceleration was not removed from the data. The velocity dispersion is 5.4 \ms and 5.8 \ms, for the pre- and after-upgrade sets, respectively.
Prior to the modelling stage, the mean radial velocity of each instrumental subset (pre- and post-upgrade) was subtracted from the corresponding measurements to improve numerical stability. Since independent systemic velocity offsets are included in all models, this normalization does not affect the inferred physical parameters.

\section{Stellar characterisation \label{sect.stellar}}

LHS~3844 is a M4.5 or M5 star \citep{vanderspek2019}, located at a distance $d=14.89\pm0.01$ pc from the Sun \citep{gaiaDR2}. \citet{vanderspek2019} determined the mass and radius of LHS~3844 to be $0.151 \pm 0.014~M_\odot$ and $0.189 \pm 0.006~R_\odot$, based on empirical relations.

The rotational period of the star was measured to be $P_\mathrm{rot}=128\pm24$ days by \citet{vanderspek2019} using 1935 photometric measurements from the MEarth-South observatory. These data were obtained over more than two and a half years, but are mostly concentrated around the observing season between March and September 2018. This value agrees with the upper limit on the rotational rate given by the rotation-activity relation of \citet{kiragastepien2007} using the X-ray upper limits of LHS~3844 from \citet{diamond-lowe2021}.

The combined ESPRESSO spectra also allow for a determination of the stellar atmospheric parameters, following the procedure by \citet{astudillo-defru17}. From the Ca II H$\&$K analysis we obtain $\log(R'_{\text{HK}})=  -5.534 \pm 0.217$. Although the low flux level at that spectral range translates into a large uncertainty, the $R'_{\text{HK}}$ provides an independent estimation of $95\pm32$ days for the stellar rotation \citep{astudillo-defru17}, consistent within $\sim1\sigma$ with the photometric determination. Given the higher precision and long baseline of the MEarth data, we adopt $P_\mathrm{rot}=128\pm24$ days as the reference value in this work.

\section{RV preliminary analysis}
\label{sect:preliminary}

We carried out a preliminary exploration of the ESPRESSO RV data independently of the transit information. This stage relied on a simplified linear modelling framework and aimed at identifying the dominant periodicities and assessing potential additional contributions to the signal. It consisted of a GLS periodogram analysis, statistical tests of additional linear covariates (stellar activity indicators and a temporal trend), and an evaluation of mild orbital eccentricity.

\subsection{Linear modelling framework} \label{sect:linealframework}

\par Given an observed RV time series, $\{t_i, v_i, \sigma_i\}$, for $i = \{1, \ldots, N\}$, where $\sigma_i$ is the velocity uncertainty based on photon noise statistics and calibration error, the velocity measurement at time $t_i$ is modeled as:

\[
v_i = m_i + \epsilon_i,
\]
where $m_i$ is the prediction of the deterministic model at time $t_i$ and $\epsilon_i$ is the error term. The variations in the RV measurements are caused by a combination of instrumental systematics, stellar activity, an additional physical acceleration, and/or the gravitational influence of orbiting companions.

As a basis for the exploratory analysis described in this section, we adopt a simplified linear model for the radial velocity time series. In this framework 
we assume that the noise terms $\epsilon_i$ are uncorrelated and normally distributed (white noise), the planet follows a circular orbit, and the orbital period is fixed. While not intended for the final parameter estimation, this approach provides a convenient and computationally efficient way to test for additional signals and assess the significance of potential model extensions.

The RV signal is modeled as:

\begin{equation}
    m_i = \gamma_{0} +  \gamma_{0}' \, \delta_{\mathrm{post}, i} + v_{\mathrm{kep}}(t_i),
    \label{eq:linear}
\end{equation}

\noindent where $\gamma_{0}$ is the global systemic velocity offset common to all observations, and $ \gamma_{0}'$ is the differential offset applied only to the post-upgrade ESPRESSO dataset. The indicator function $\delta_{\mathrm{post}, i}$ equals 1 if the $i$-th observation belongs to the post-upgrade dataset, and 0 otherwise. Under the assumption of circular orbit with fixed period $P$, $v_\mathrm{kep}$ is given by:

\[
v_\mathrm{kep}(t) = \beta_0 \sin\left(\frac{2\pi t}{P}\right) + \beta_1 \cos\left(\frac{2\pi t}{P}\right),
\]

\noindent from which the semi-amplitude $K$ of the radial velocity variation can be expressed as $K^2 = \beta_0^2 + \beta_1^2$. 

The initial assumption of circularity is justified by the ultra-short orbital period of the planet, which leads to strong tidal interactions that damp orbital eccentricity and drive the system toward a nearly circular configuration \citep[e.g.][]{hut1980}, but we lift this constraint  below.

\subsection{Periodogram analysis}
\label{sect.periodogram}

\begin{figure}
    \centering
    \includegraphics[width=\columnwidth]{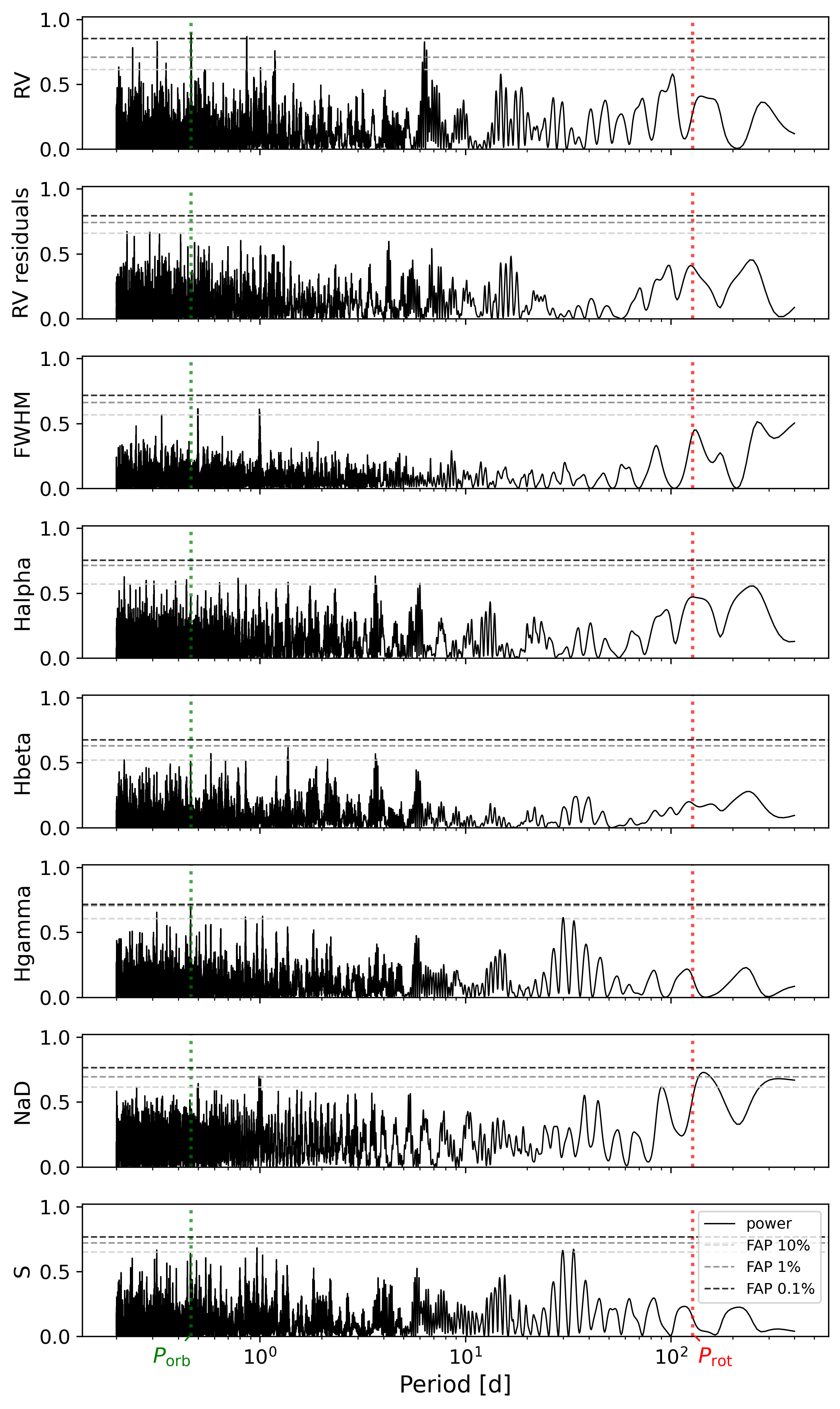}
    \caption{GLS Periodograms \citep{zechmeisterkurster2009} for the ESPRESSO radial velocities, their residuals after removing the best fit model for a circular orbit, and several stellar activity indicators measured for LHS~3844. Vertical dashed lines mark the transit period $P = 0.46$~days (green) and the rotational period of the star $P_{\text{rot}} = 128$~days (red) \citep{vanderspek2019}. The 10, 1 and 0.1 per cent False Alarm Probabilities (FAP) are shown as dashed horizontal lines.}

    \label{fig:GLSrv}
\end{figure}

 In Fig.~\ref{fig:GLSrv} we present the Generalised Lomb Scargle (GLS) periodogram \citep{zechmeisterkurster2009} of the velocities and stellar activity indicators. We have allowed for a frequency-dependent offset between the measurements acquired before and after the instrumental upgrade of June 2019 for each time series, as shown in Eq. \ref{eq:linear}. 
\par In the RV periodogram (topmost panel) we detect a peak at $P_{\text{max}} = 0.4628 \pm 0.0007$ d above the 0.1 per cent False Alarm Probability which lies within $0.1 \sigma$ of the transit-determined period. A second peak above the $0.1$ $\%$ FAP level appears at 0.86 d, corresponding to a first armonic alias of the 0.46 d signal. After removing the best-fit sinusoidal model for the stellar RV variation at that period, the periodogram of the RV residuals (second panel) shows no further significant peaks. Within the scope of this simple model, we therefore find no evidence for an additional planetary signal in this analysis.

With the exception of the Na I D index (penultimate panel), no other activity index exhibits significant periodicity at the rotational rate of the star. Also, no clearly significant peaks are seen around the period of the planetary companion, but the periodogram of the H$\gamma$ index shows a peak with an amplitude at the level of 0.1$\%$.  

\subsection{Testing additional covariates}\label{sect.linear}

To further examine the effects of stellar activity on the RV data, we tested if the RV curves of our adopted models described in Section~\ref{sect.rvanalysis} should include additional linear terms associated with stellar activity indicators or a velocity trend over time. For this we built simple linear models in which the radial velocity was modeled as a function of each activity indicator separately, in addition to the base components (instrumental offsets). Each extended model was compared to the corresponding base model using F-tests. In all cases, the resulting $p$-values exceeded 0.01. In addition, neither the Akaike nor the Bayesian Information Criteria (AIC, BIC, respectively) indicated any significant improvement.

We then incorporated a circular orbit signal with a fixed period corresponding to the highest peak in the GLS periodogram into each model and repeated the comparisons against the circular-orbit-only model. Once again, none of the activity indicators yielded $p$-values below 0.01, and the AIC and BIC values similarly showed no evidence for an improved fit. This shows that, at least within the domain of the tested linear model, it is not necessary to incorporate linear terms with these covariates to explain the RV data. 

\subsection{Testing mild orbital eccentricity}

We also explored the possibility that the planetary orbit may exhibit a small but non-zero eccentricity. To keep the model linear, we used the first order Taylor expansion in eccentircity of the Keplerian model, which led to the inclusion of two additional periodic terms at the first harmonic of the orbital period (see Appendix \ref{appendix_orbital_model_ecc}). 
This approach allows us to test for eccentricity effects with all the machinery of linear models.

We compared the extended mildly eccentric model (with the additional harmonic terms) to the simpler circular orbit model using F-tests and information criteria (AIC/BIC). The results showed no statistically significant improvement: all $p$-values were greater than 0.01, and both AIC and BIC values favoured the circular model. We therefore conclude that a small orbital eccentricity is not supported by the data and adopt the simpler circular model in the subsequent analysis.

\section{Radial velocity modelling}
\label{sect.rvanalysis}

We now move beyond the simplified linear framework adopted in the preliminary analysis and develop a comprehensive Bayesian model for the ESPRESSO radial velocities. This framework simultaneously accounts for correlated noise through Gaussian Process regression, instrumental systematics, possible long-term velocity drifts, and informative priors on the orbital parameters. All model parameters and hyperparameters are treated as free and inferred jointly from the data.

\subsection{Deterministic model}

The deterministic component of the radial velocity signal consists of a Keplerian contribution and instrumental offsets, optionally supplemented by linear or quadratic velocity drifts. The component of the model not associated with the planetary signal is expressed as:

\[
\gamma_0 + \gamma_0'\,\delta_{\text{POST}, i} + \gamma_1 (t_i-t_\text{ref})  + \gamma_2 (t_i-t_\text{ref})^2,
\]

\noindent where $\gamma_1$ and $\gamma_2$ correspond to the linear and quadratic drift terms, respectively. We adopt $t_\mathrm{ref} = 2458578.66$ d (BJD), chosen close to the midpoint of the observations in order to reduce parameter correlations.

\noindent We define three nested versions of this component: a constant-only model ($d_0$), a model including a linear drift term ($d_1$), and a model incorporating both linear and quadratic drift terms ($d_2$).

To incorporate prior constraints from transit photometry, we express the circular Keplerian signal in terms of the semi-amplitude $K$, orbital period $P$, and phase $\phi$:

\[
v_\mathrm{kep}(t_i) = K \sin\left( \frac{2\pi}{P} t_i + \phi \right),
\]

\noindent where, for a circular orbit, the phase relates to the transit mid-time as $ \phi = \pi - \frac{2\pi}{P} t_0$. This choice allows external information to be directly imposed through $\phi$ and $P$.

\subsection{Error model and Gaussian Process regression}

\par The error terms $\boldsymbol{\epsilon} = (\epsilon_1, \ldots, \epsilon_N)^\top$ account for both uncorrelated (white noise) and correlated noise sources. The uncorrelated noise component includes the measurement uncertainties $\sigma_i$ and an additional jitter term $\sigma_{Jt}$ for each instrument. Correlated noise is modeled using a Gaussian Process (GP) with a covariance computed from a kernel function $k(t_i, t_j)$, which here will be stationary and depend exclusively on the time between observations $|t_i - t_j|$. 

\par We explore the dependence of the inferred planetary mass with the following types of kernels: white noise (WN), squared-exponential (SE), periodic (Per), and quasi-periodic (QP) kernels. The kernel functions depend on the amplitude $A$, a characteristic timescale $\tau$, a periodic timescale $\mathcal{P}$, and a smoothness parameter $\lambda$, with functional forms given by:

\begin{align*}
k_\mathrm{WN}(t_i, t_j) =& A^2 \delta_{ij}, \\
k_\mathrm{SE}(t_i, t_j) =& A^2 \exp\left(-\frac{(t_i - t_j)^2}{2\tau^2}\right), \\
k_\mathrm{Per}(t_i, t_j) =& A^2 \exp\left(-\frac{2\sin^2\left(\frac{\pi |t_i - t_j|}{\mathcal{P}}\right)}{\lambda^2}\right), \\
k_\mathrm{QP}(t_i, t_j) =& A^2 \exp\left(-\frac{(t_i - t_j)^2}{2\tau^2}\right)
                          \exp\left(-\frac{2\sin^2\left(\frac{\pi |t_i - t_j|}{\mathcal{P}}\right)}{\lambda^2}\right).
\end{align*}

\par
To characterize a broad range of temporal correlations present in RV time series, we test the following five kernel compositions, using the fact that the sum of valid kernel functions remains a valid kernel function.

\begin{enumerate}
    \item WN
    \item WN + SE
    \item WN + QP
    \item WN + SE + QP
    \item WN + SE + Per
\end{enumerate}

\noindent
In all models we include both an instrumental jitter term for each version of ESPRESSO and a common white-noise term. The latter is intended to capture stellar variability that is equally present in both instruments but not adequately sampled by our data (e.g.~granulation effects).

Finally, the covariance matrix for the error terms is then given by:
\begin{equation}
    \Sigma_{ij} = k(t_i, t_j) + \delta_{ij}\left(\sigma_i^2 + \sigma_{Jt}^2\right)
\label{eq:covariance}
\end{equation}

\noindent where $k(t_i, t_j)$ is the covariance function constructed using one of the kernel combinations above.

\subsection{Model priors and likelihood}\label{sec:priorslikelihood}

We explored five alternative GP kernel prescriptions combined with three nested versions of the deterministic drift component ($d_0$, $d_1$, $d_2$), resulting in a total of 15 models. All parameters and hyperparameters were assigned the prior distributions summarized in Table~\ref{table:priors}, which were consistently applied across all models.

\begin{table}
\caption{
Prior distributions adopted for the parameters in the radial velocity modelling.}
\small
\begin{tabular}{l l c}
\hline\hline                 
\noalign{\smallskip}

	\textbf{Parameter} & \textbf{Units} & \textbf{Prior}\\
\noalign{\smallskip}

\multicolumn{3}{c}{\textbf{Linear parameters}} \\
\hline
\noalign{\smallskip}
General offset $\gamma_0$ & [m s$^{-1}$] & $\mathcal{N}(0,10)$ \\
ESPRESSO-post offset $\gamma_0'$& [m s$^{-1}$] & $\mathcal{N}(0,10)$ \\
RV semi-amplitude $K$ & [m s$^{-1}$] & $\mathcal{N}(0,10)$ \\
Linear trend $\gamma_1$ & [m s$^{-1}$d$^{-1}$] & $\mathcal{N}(0,10)$ \\
Quadratic trend $\gamma_2$ & [m s$^{-1}$d$^{-2}$] & $\mathcal{N}(0,10)$ \\

\noalign{\smallskip}
\multicolumn{3}{c}{\textbf{Non-linear parameters}} \\
\hline
\noalign{\smallskip}
Jitter (pre), $\sigma_{Jt}^{\text{pre}}$ & [m s$^{-1}$] & $\ln \mathcal{U}(10^{-3}, 10)$\\
Jitter (post), $\sigma_{Jt}^{\text{post}}$ & [m s$^{-1}$] & $\ln \mathcal{U}(10^{-3}, 10)$\\
Orbital period, $P$ & [d] & $\ln \mathcal{N}(P_{\text{t}}, \sigma_{P_\text{t}})$\\
Transit mid-time, $t_0$ & [d] & $\mathcal{N}(t_{0\text{t}}, \sigma_{t_{0\text{t}}})$\\

\noalign{\smallskip}
\multicolumn{3}{c}{\textbf{GP hyperparameters}} \\
\hline
\noalign{\smallskip}

\multicolumn{3}{c}{White noise} 		\\
\hline
\noalign{\smallskip}
White-noise amplitude, $A_\mathrm{white}$ & [m s$^{-1}$] & $\ln \mathcal{U}(10^{-3}, 10)$\\

\noalign{\smallskip}
\multicolumn{3}{c}{SE kernel} 		\\
\hline
\noalign{\smallskip}
Covariance amplitude, $A_\mathrm{SE}$ & [m s$^{-1}$] & $\ln \mathcal{U}(10^{-3}, 10)$\\
Decay timescale, $\tau_\mathrm{SE}$ & [d] & $\ln \mathcal{U}(3 P_\text{rot}, \ln{(5)}/2)$\\

\noalign{\smallskip}
\multicolumn{3}{c}{Per kernel} 		\\
\hline
\noalign{\smallskip}
Covariance amplitude, $A_\mathrm{per}$ & [m s$^{-1}$] & $\ln \mathcal{U}(10^{-3}, 10)$\\
Periodicity, $\mathcal{P}_\mathrm{per}$ & [d] & $\ln \mathcal{N}(P_{\text{rot}}, \sigma_{P_\text{rot}}/P_{\text{rot}})$\\

Smoothness parameter, $\lambda_\mathrm{per}$ & [-] & $\ln \mathcal{U}(0.5, 2)$\\

\noalign{\smallskip}
\multicolumn{3}{c}{QP kernel} 		\\
\hline
\noalign{\smallskip}
Covariance amplitude, $A_\mathrm{QP}$ & [m s$^{-1}$] & $\ln \mathcal{U}(10^{-12}, 10)$\\
Decay timescale, $\tau_\mathrm{QP}$ & [d] & $\ln \mathcal{U}(3 P_\text{rot}, \ln{(5)}/2)$\\
Periodicity, $\mathcal{P}_\mathrm{QP}$ & [d] & $\ln \mathcal{N}(P_{\text{rot}}, \sigma_{P_\text{rot}}/P_{\text{rot}})$\\
Smoothness parameter, $\lambda_\mathrm{QP}$ & [-] & $\ln \mathcal{U}(0.5, 2)$\\

\hline
\end{tabular}

{\captionsetup{font=small}
\caption*{\textbf{Notes.} $\mathcal{N}(\mu, \sigma)$ and $\ln\mathcal{N}(\mu, \sigma)$ denote a normal and log-normal distributions, respectively, with mean $\mu$ and standard deviation $\sigma$. \\
$\ln\mathcal{U}(a,b)$ is a log-uniform distribution between $a$ and $b$.\\
$P_{\mathrm{t}}$, $\sigma_{P_{\mathrm{t}}}$, $t_{0\text{t}}$ and $\sigma_{t_{0\text{t}}}$ correspond to the orbital period and transit-mid time obtained from transit observations and their associated uncertainty \citep{kokori2023}.
$P_{\text{rot}}$ and $\sigma_{P_{\mathrm{rot}}}$ correspond to the stellar rotation period and its uncertainty reported in Sect \ref{sect.stellar}.}
}
\label{table:priors}
\end{table}

Informative priors were adopted for the orbital period $P$ and transit mid-time $t_0$, based on the results of the transit analysis \citep{kokori2023}. The reported transit epoch was propagated to the temporal midpoint of the radial velocity observations by shifting it by an integer number of orbital periods. The corresponding uncertainty was updated by propagating the period error accordingly. This choice ensures that the phase prior is referenced to the time span of the RV data, reducing parameter correlations and improving numerical stability. Through the phase relation defined in the previous subsection, the prior on $t_0$ is directly propagated into the radial velocity model. For kernels including a quasi-periodic component, the periodic timescale was assigned a prior informed by the stellar rotation period measured in Sect.~\ref{sect.stellar}.

Linear parameters $\boldsymbol{\beta} = \{\gamma_0,\gamma_0',\gamma_1,\gamma_2,K\}$ were assigned multivariate normal priors,
$\boldsymbol{\beta} \sim \mathcal{N}(\boldsymbol{\mu}_\beta, B)$. 
Under this assumption, the marginal likelihood obtained after analytically integrating over the linear parameters (see \citet{rasmussen}, Sects.~2.7.1 and 5.4.1) can be written as

\[
\ln p(\mathbf{v} \mid \theta)
=
-\frac{1}{2}
\left[
\left(\boldsymbol{m} - \mathbf{v}\right)^{\!\top}
K_v^{-1}
\left(\boldsymbol{m} - \mathbf{v}\right)
+ \ln\left|K_v\right|
+ n \ln(2\pi)
\right],
\]

\noindent where $\mathbf{v}$ is the vector of observed radial velocities, $\boldsymbol{m} = X\boldsymbol{\mu}_\beta$ corresponds to the model evaluated at the prior mean of the linear parameters, $n$ denotes the number of observations and $K_v = \Sigma + X B X^\top$, with $X$ being the design matrix and $\Sigma$ the covariance matrix given by Eq. \ref{eq:covariance}.
The vector of non-linear parameters $\boldsymbol{\theta}$ includes the orbital period $P$, transit mid-time $t_0$, instrumental jitter terms, and the hyperparameters of the GP kernels. These parameters enter the likelihood through both the deterministic and stochastic components of the model.

\subsection{Posterior distribution sampling}
\label{sect.mcmc}

The posterior distributions of each model’s parameters were explored using a Markov Chain Monte Carlo (MCMC) algorithm, sampling only the non-linear parameter vector $\boldsymbol{\theta}$ through the marginal likelihood defined in Sect.~\ref{sec:priorslikelihood}, where the linear parameters have been analytically integrated out. We used \texttt{exoplanet} to first obtain the maximum a posteriori (MAP) solution for each model by maximizing the log-probability. The resulting parameter values were then used to initialize the \texttt{PyMC3} sampler, which draws samples from the posterior using the No-U-Turn Sampler (NUTS), a variant of Hamiltonian Monte Carlo. For each of the 15 models, we ran 10 independent chains. We allowed for 15\,000 tuning steps, which were discarded, followed by 30\,000 posterior samples per chain. Convergence was assessed using the rank-normalized split-$\hat{R}$ statistic \citep{rhat}. For all models, the vast majority of parameters exhibit $\hat{R} < 1.01$, indicating satisfactory mixing and convergence.

For models including a quadratic drift component ($d_2$), a small number (1–3) of chains were discarded due to persistent divergences during sampling. In these cases, the remaining chains showed stable behaviour. Additionally, in the $d_2$ models, the white-noise hyperparameters (instrumental jitter terms and the common white-noise amplitude) exhibited slightly elevated $\hat{R}$ values, reaching up to 1.02. This behaviour is likely due to increased posterior correlations between curvature terms and noise amplitudes, which can introduce mild sampling inefficiencies without significantly affecting the inferred planetary parameters. All other parameters in these models satisfy $\hat{R} < 1.01$.

Posterior samples of the linear parameters $\boldsymbol{\beta}$ were subsequently obtained by drawing from their conditional multivariate normal distribution,
$p(\boldsymbol{\beta} \mid \mathbf{v}, \boldsymbol{\theta})$, evaluated at each posterior sample of $\boldsymbol{\theta}$. In this way, full posterior samples for all model parameters were recovered while retaining the computational advantages of marginalizing over the linear components during the MCMC exploration.

To compute the marginal likelihood of each model, we employed the estimator proposed by \citet{Perrakis_2014}, first implemented in an astronomical context by \citet{diaz2016a}. The estimator relies on importance sampling using draws from the marginal posterior distribution of the nonlinear parameters. The estimator relies on importance sampling using draws from the product of the marginal posterior distributions of the nonlinear parameters. To approximate independent samples from this distribution, the MCMC chains were randomly permuted independently for each parameter prior to evaluation, thereby breaking the correlations present in the joint posterior samples. For each model, the log-marginal likelihood was computed from subsamples of increasing size (from 500 up to 60,000 draws), with each configuration repeated 10 times using independent resamplings. We adopt as the final estimate the mean value obtained with 60,000 samples. In all cases, the estimator showed clear convergence as the number of samples increased, with the dispersion decreasing monotonically, indicating numerical stability of the evidence estimates.

\section{Results and discussion \label{sect.discussion}}

\subsection{Model comparison}
\label{sect:modelcomparison}

\begin{figure}
    \centering
    \includegraphics[width=\columnwidth]{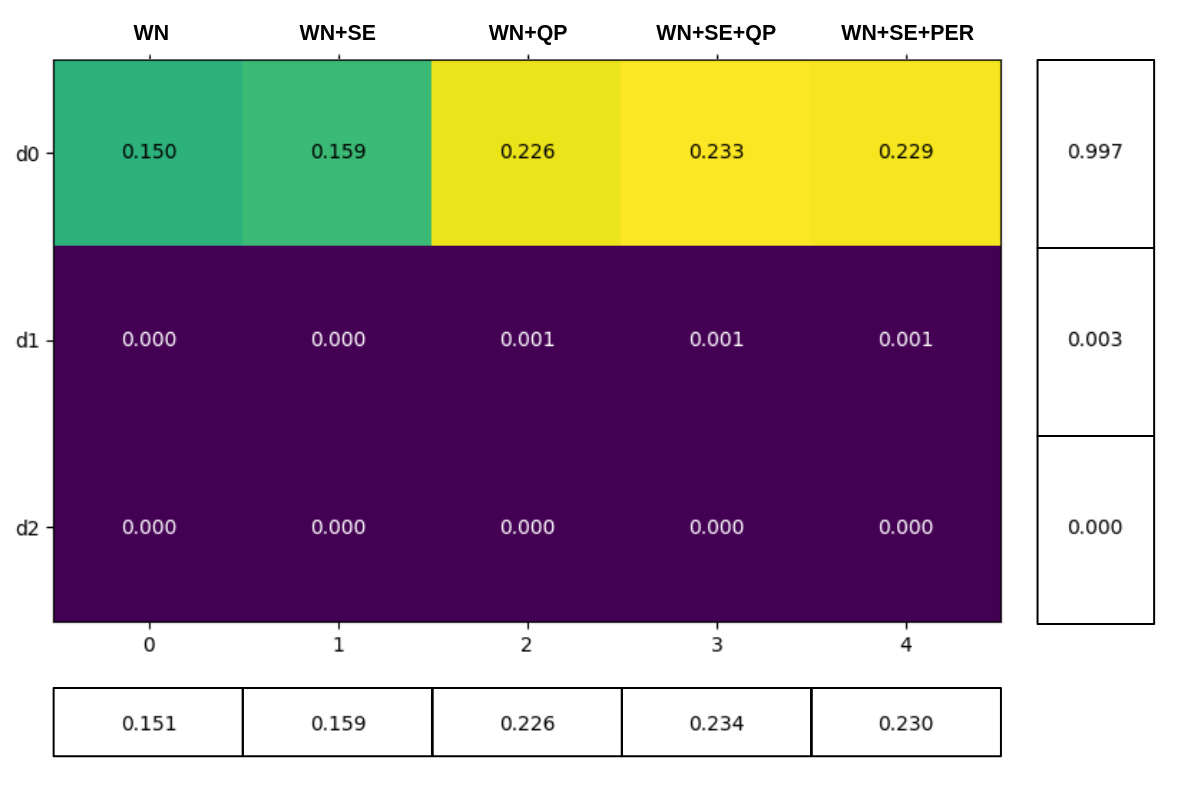}
    \caption{Posterior probability matrix for the 15 evaluated models, spanning all combinations of three secular acceleration models and five kernels. The numbers inside the cells indicate the normalized posterior probability of each model. The rightmost column and bottom row show the marginalized probabilities over drift and kernel types, respectively.}
    \label{fig:evidence_matrix}
\end{figure}

Using the marginal likelihood estimates described in Sect.~\ref{sect.mcmc}, we compare the full set of models explored in this work. Figure~\ref{fig:evidence_matrix} presents the posterior probability of each model within the explored set of 15 kernel–drift combinations. These probabilities were obtained by normalizing the marginal likelihoods under the assumption of equal prior probability for all models in the considered model space. The last row and column show the probabilities marginalized over the drift and kernel dimensions, respectively. Models without drift ($d_0$) consistently dominate, regardless of the chosen kernel. This outcome is consistent with the preliminary linear modelling analysis (Sect.~\ref{sect.linear}), where adding temporal drift terms did not produce statistically significant improvements based on F-tests or changes in AIC/BIC. Among kernel configurations, those including periodic or quasi-periodic components are favoured over simpler white-noise or squared-exponential kernels.

\begin{figure*}
    \centering
    \includegraphics[width=\textwidth]{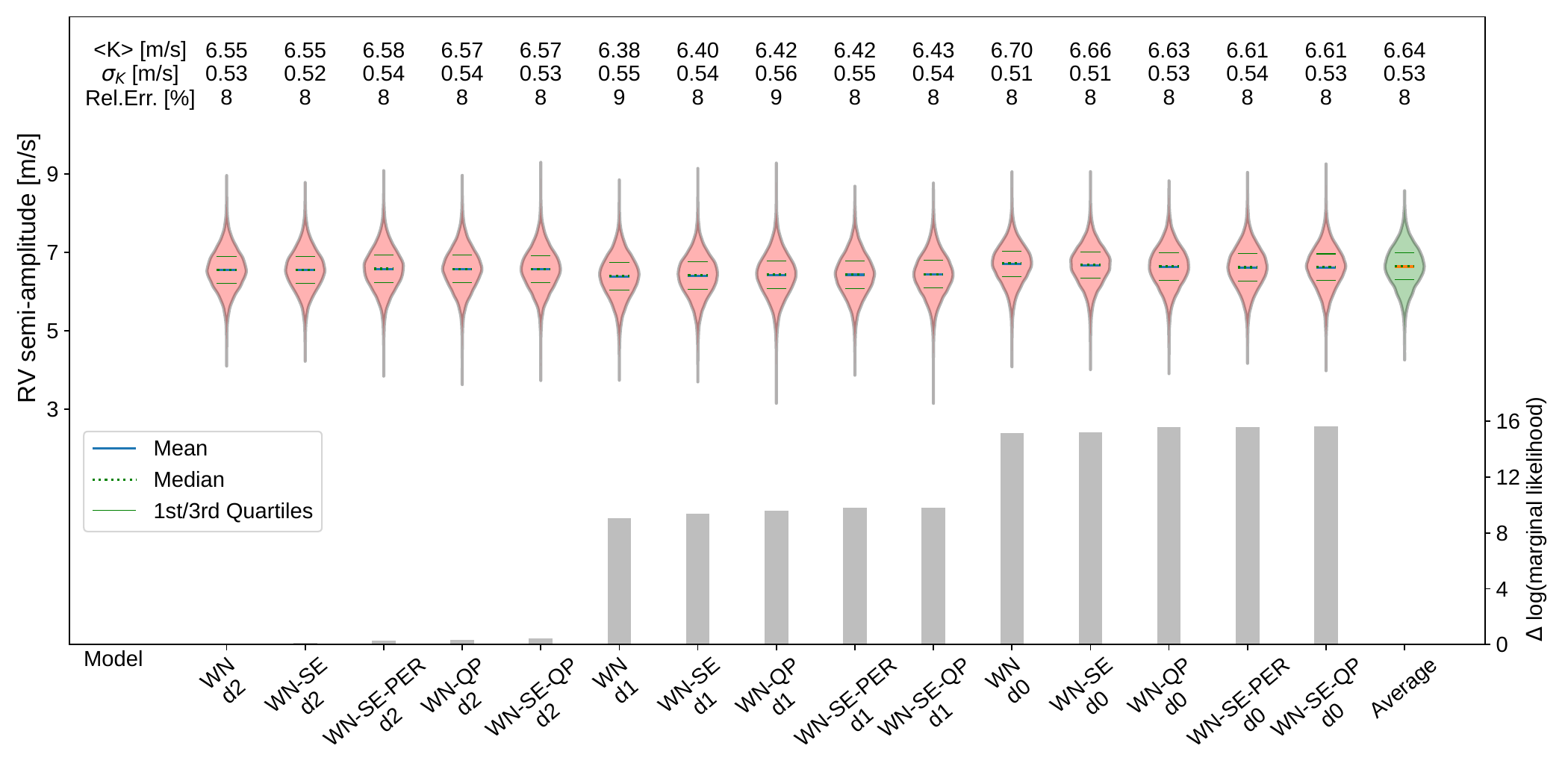}
    \caption{Comparison of the inferred RV semi-amplitude ($K$) across different models, ordered by increasing marginal likelihood. Each violin plot represents the posterior distribution of $K$ for a given model, with the mean (solid line), median (dotted line), and 1st/3rd quartiles (green). The rightmost violin corresponds to the weighted average model. The overlaid bar plot (grey) shows the relative log-likelihood difference for each model. The numerical annotations indicate the mean, standard deviation, and relative uncertainty of $K$ for each model.}
    \label{fig:models_semiamplitude}
\end{figure*}

\subsection{Inferred semi-amplitude and dominant models}
\label{sect.K}

Figure~\ref{fig:models_semiamplitude} shows the distribution of the RV semi-amplitude ($K$) as inferred from the full set of models explored in this work. Each violin plot represents the posterior distribution of $K$ for a specific combination of GP kernel and drift term, while the rightmost violin shows the marginal distribution across all models, weighted by their marginal likelihood. 
Despite substantial differences in the adopted noise prescriptions (from simple white noise to multi-component quasi-periodic kernels) and in the inclusion or exclusion of temporal drift terms, all models yield very similar posterior distributions for $K$. In practice, the inferred semi-amplitudes consistently fall in the range $6$–$7$~m\,s$^{-1}$, indicating that the estimate is largely insensitive to the adopted noise model.

As an additional robustness check, we repeated the analysis using a Maximum Likelihood Type-II (ML-II) approach, described in Appendix~\ref{appendix:ml2}. Although this method fixes the orbital period and epoch and optimises the marginal likelihood with respect to the hyperparameters rather than sampling them, the resulting estimates of $K$ are fully consistent with those obtained from the full Bayesian MCMC framework. This agreement across fundamentally different inference schemes further supports the stability of the inferred semi-amplitude.

The posterior predictive distributions for the highest-probability model (\texttt{WN+SE+QP:}  $d_0$) are shown in Figure~\ref{fig:phase_folded_rv}, which presents the phase-folded orbital component and in Figure~\ref{fig:GP_fit} in Appendix~\ref{appendix_plots_tables}, which shows the GP predictive fits to the residuals.

\subsection{Inferred planetary properties of LHS 3844\,b}
\label{sect.inferredproperties}

\begin{table*}[ht]
    \renewcommand{\arraystretch}{1.4}
    \centering
    \caption{Planetary and orbital parameters of LHS 3844\,b. }
	\label{tab:params_LHS3844}
	\begin{tabular}{lccc}
	\hline
	\textbf{Parameter} & \textbf{(Unit)} & \textbf{Value} & \textbf{Source}\\
	\hline
	Period $P$ & (days) & $0.4629296 \pm 0.0000003$ & This work $^{\dagger}$\\
	Transit mid-time $t_0$ & (BJD-2457000) & $1551.1726 \pm 0.0003$ & This work $^{\dagger}$\\
    Radial velocity semi-amplitude $K$ & (m\,s$^{-1}$) & $6.65 ^{+0.49}_{-0.53}$ & This work\\
    Minimum mass $M_p \sin{i}$ & (M$_{\oplus}$) & $2.27 \pm 0.23$ & This work\\
    Eccentricity $e$ & -- & $0.0$ (fixed) & This work\\

    \hline
    Mass $M_p$ & (M$_{\oplus}$) & $2.27 \pm 0.23$ & This work + \citep{vanderspek2019}\\
    Bulk density $\rho$ & (g\,cm$^{-3}$) & $5.67 ^{+0.65}_{-0.61}$ & This work + \citep{vanderspek2019}\\
    Semi-major axis $a$ & (AU) & $0.00624 \pm 0.00019$ &This work + \citep{vanderspek2019}\\
    $T_{\mathrm{eq}}$ (albedo $\zeta=0.0$) & (K) & $806 \pm 27$ & This work + \citep{vanderspek2019}\\
    $T_{\mathrm{eq}}$ ($\zeta=0.3$) & (K) & $737 \pm 25$ & This work + \citep{vanderspek2019}\\
    $T_{\mathrm{eq}}$ ($\zeta=0.5$) & (K) & $678 \pm 23$ & This work + \citep{vanderspek2019}\\
    \hline
    Radius $R_p$ & (R$_{\oplus}$) & $1.303 \pm 0.022$ & \citet{vanderspek2019}\\
    Radius ratio $R_p/R_\star$ & -- & $0.0635 \pm 0.0009$ & \citet{vanderspek2019}\\
    Inclination $i$ & (deg) & $88.50 \pm 0.51$ & \citet{vanderspek2019}\\
    Impact parameter $b$ & -- & $0.186 \pm 0.064$ & \citet{vanderspek2019}\\
	\hline
	\end{tabular}
    {\captionsetup{font=small}
    \caption*{\textbf{Notes.}  The values given are the medians, 16th and 84th percentiles of the posterior distributions, derived by marginalizing over all models using ancestral sampling. When indicated, values combine the results of this work with the transit analysis of \citet{vanderspek2019}, or are taken directly from that work.\\ $^{\dagger}$ Strongly constrained by the transit-based prior \citep{kokori2023}. }}

\end{table*}

Using the posterior samples described in Sect.~\ref{sect.mcmc}, we derived the physical parameters of LHS 3844\,b by marginalizing over the full set of models through ancestral sampling, adopting the marginal likelihood as the weighting factor in the model averaging. The reported values correspond to the median and the 16th and 84th percentiles of the resulting marginalized posterior distributions. A summary of the planetary parameters is given in Table~\ref{tab:params_LHS3844}, including the estimates obtained in this work, values derived by combining our results with the transit analysis of \citet{vanderspek2019} and \citet{kokori2023}, and those reported directly from the transit fit.

The orbital period is determined to be $P = 0.4629296 \pm 0.0000003$ days, consistent with the prior based on transit observations, as expected given that the RV data alone carry less information on the planet's periodicity than the transit measurements.  From the period we also derive the orbital semi-major axis $a$ and the equilibrium temperatures for different albedo assumptions, which are necessarily compatible with the transit-based results. 
The short semi-major axis may suggest the possibility of star--planet interactions \citep{shkolnik2003, shkolnik2008}. Although we tentatively detect variations in the H$\gamma$ index at the orbital period of the planet (Sect.~\ref{sect.periodogram}), such behaviour is not observed in other indicators, including those tracing regions higher in the stellar atmosphere. The available evidence therefore remains too limited to draw firm conclusions at this stage.

The RV semi-amplitude obtained in this work is $K = 6.65 ^{+0.49}_{-0.53}$~m\,s$^{-1}$, yielding a minimum mass of $M_p \sin i = 2.27 \pm 0.23$~M$_\oplus$. The quoted uncertainty includes contributions from the errors in $K$ and $P$ (this work) and the stellar mass (from \citealt{vanderspek2019}). This corresponds to a relative uncertainty of roughly 10\%. Using the inclination measured from the transit geometry ($i = 88.5^\circ$), the absolute mass is effectively indistinguishable from the minimum mass, as the difference is much smaller than the associated uncertainty ($0.02$ $\sigma$). Finally, the bulk density is found to be $\rho = 5.67 ^{+0.65}_{-0.61}$~g\,cm$^{-3}$ by combining our RV-based mass with the planetary radius derived from transit observations. This value indicates a composition consistent with a rocky super-Earth.

\begin{figure}
    \centering
    \includegraphics[width=\columnwidth]{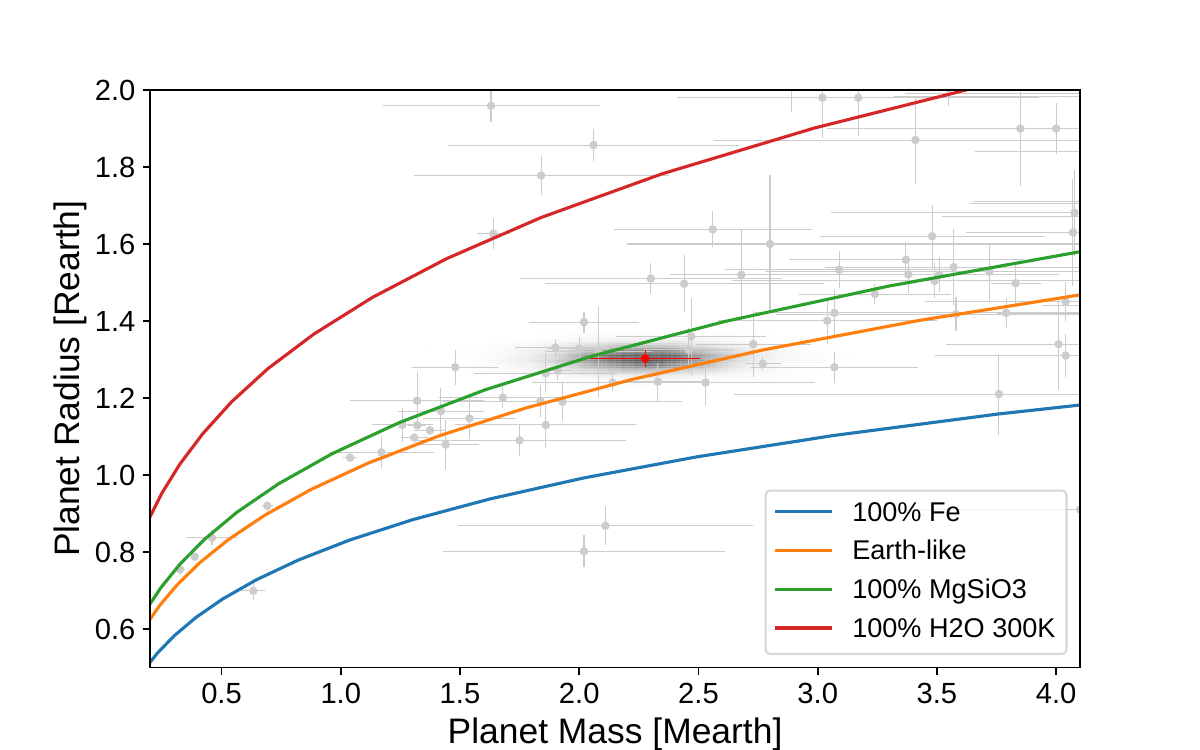}
    \caption{Mass-radius diagram for small exoplanets ($M_p < 5\,M_\oplus$) with precisely measured masses ($\sigma_M / M < 0.3$). Gray points represent the observed exoplanets with error bars, while the red point marks LHS~3844~b. The shaded background corresponds to a kernel density estimation (KDE) of the joint posterior distribution of mass and radius for LHS~3844~b.  Solid lines indicate theoretical mass-radius relations for different compositions from \citet{Zeng2016}. Known planets were sourced from the NASA Exoplanet Archive (\url{https://exoplanetarchive.ipac.caltech.edu/}) on 26 February 2026.}

    \label{fig:MR}
\end{figure}

In Figure~\ref{fig:MR}, we place LHS 3844\,b in the context of the known population of small exoplanets with well-constrained masses ($\sigma_M / M < 0.3$). As of February 2026, among the 6107 confirmed exoplanets, only 1072 (17.5\%) have mass determinations with a relative uncertainty better than 10\%, the precision we obtain here for LHS 3844\,b.
The planet lies close to the mass–radius relation expected for terrestrial iron–silicate compositions, but its peak posterior probability favours a slightly lower bulk density than an Earth-like composition, consistent with a rocky interior between the Earth’s composition and a pure silicate (100\% mantle) model.
Taken together, our results confirm LHS 3844\,b as an ultra-short-period, rocky exoplanet with well-constrained mass and density, likely of terrestrial composition.

\begin{figure}
    \centering
    \includegraphics[width=\columnwidth]{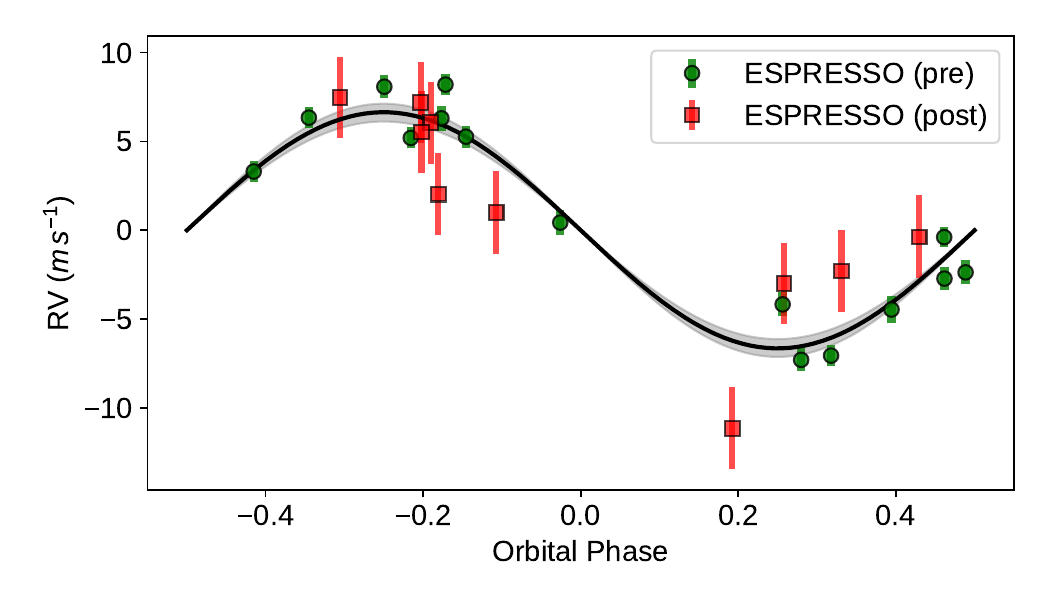}
    \caption{Phase-folded RV curve of \object{LHS~3844}, based on the best-fit model with the highest marginal likelihood: White + SE + QP Kernel with no drift ($d_0$). The data points correspond to ESPRESSO observations obtained before (green circles) and after (red squares) the instrumental upgrade. The data were corrected by subtracting the per-instrument velocity offsets but not the noise component of the GP model. Phase folding was performed using the median of the posterior distribution for the orbital period. Error bars represent the quadrature sum of the reported uncertainties and the empirically inferred per-instrument jitter. The black curve shows the median orbital model (circular orbit), while the gray shaded region represents the 68\% credible interval derived from posterior samples. }
    \label{fig:phase_folded_rv}
\end{figure}

\subsection{Linear and non-linear parameters}
\label{sect.linparam}

\par Summaries of the inferred posterior distributions for both nonlinear and linear model parameters are provided in Appendix~\ref{appendix_plots_tables}. Table~\ref{tab:nonlinear_parameters_kernel} reports the marginalized means and standard deviations of the nonlinear hyperparameters grouped by kernel type, while Table~\ref{tab:linear_parameters_drift} presents the corresponding results for linear parameters, grouped by acceleration model. These tables offer a detailed overview of model dependencies and provide additional context for interpreting the model selection based on marginal likelihood computation, as discussed above.

\par We find that the model requires distinct jitter terms for the pre- and post-upgrade data sets. The posterior distributions of $\ln \sigma_{J}$ (pre) cluster around $-2.0$ to $-2.3$ (i.e., $\sim 0.1$~m\,s$^{-1}$), whereas $\ln \sigma_{J}$ (post) spans $\sim -0.2$ to $0.7$ (i.e., $\sim 1$–$2$~m\,s$^{-1}$), indicating a substantially larger excess variance in the post-upgrade data. This effect is most pronounced for kernels without a periodic component (WN and WN+SE), where the post-upgrade jitter dominates the variance budget. In contrast, the amplitudes of the time-correlated GP components (SE or QP/Per) are small compared to the post-upgrade jitter and typically comparable to or below the white-noise amplitude. As a result, the correlation hyperparameters ($\tau$, $\lambda$, $\mathcal{P}_{\mathrm{GP}}$) remain weakly constrained and largely reflect their priors, consistent with the limited data set. 

\begin{figure*}
    \centering
    \includegraphics[width=\textwidth]{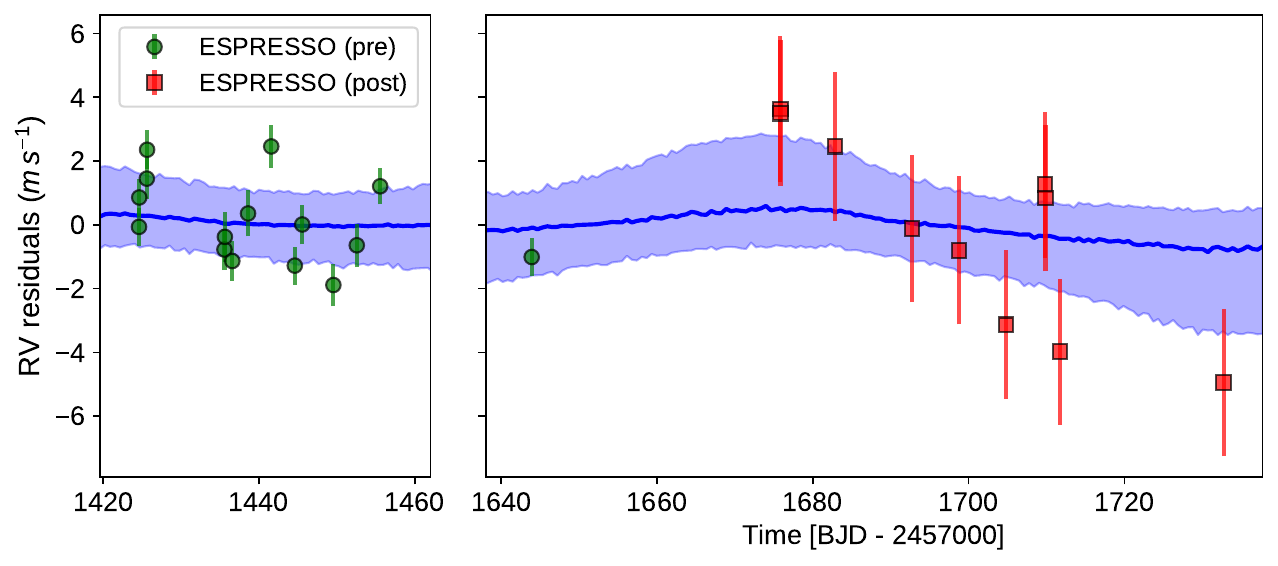}
    \caption{RV residuals of \object{LHS~3844} after subtracting the per-instrument offsets and the orbital component, sampled from the posterior distribution of the model with highest marginal likelihood: WN+SE+QP kernel without drift terms ($d_0$). The blue curve represents the median of the Gaussian Process (GP) predictive distribution, computed over the full posterior of the GP hyperparameters and marginalized over the linear parameters and orbital period. The shaded region denotes the 68\% confidence interval. Data points correspond to ESPRESSO observations taken before (green circles) and after (red squares) the instrumental upgrade. Error bars include both the reported measurement uncertainties and the empirically derived per-instrument jitter, added in quadrature.}
    \label{fig:GP_fit}
\end{figure*}

Although periodic kernels introduce a mild preference for a characteristic timescale (see Fig.~\ref{fig:GP_fit}), the correlated component is weakly required. Overall, the variance is dominated by instrument-dependent jitter, and the RV signal is driven primarily by the planetary component, with no strong evidence for coherent stellar activity.

Regarding drift terms, the coefficients of the linear ($\gamma_1$) and quadratic ($\gamma_2$) trends show at most marginal deviations from zero (at the $\lesssim 2\sigma$ level), and their inclusion leads to a significant drop in marginal likelihood (see Fig.~\ref{fig:evidence_matrix}). This reinforces the conclusion from the linear models and the Bayesian model comparison that long-term trends are not supported by the data.

\subsection{Second periodic signal} \label{sect:secondsignal}

\begin{figure}
    \centering
    \includegraphics[width=\columnwidth]{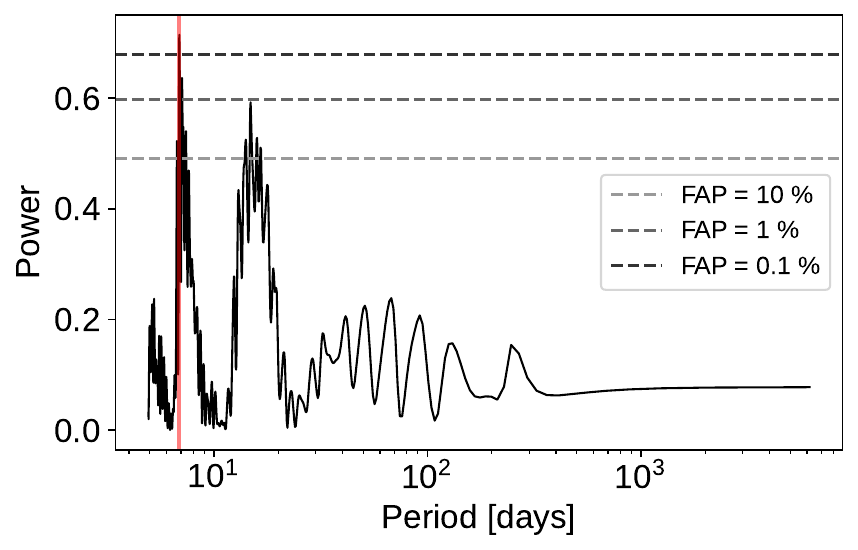}
    \caption{Generalized Lomb--Scargle periodogram of the radial-velocity residuals after subtracting the median posterior predictive deterministic component (instrumental offsets and orbital signal) of the model with highest marginal likelihood (WN+SE+QP kernel without drift terms, $d_0$). Horizontal dashed lines indicate the false-alarm probability (FAP) levels of 10\%, 1\%, and 0.1\%. The vertical red line marks the strongest peak in the periodogram at $P = 6.886$ days.}

    \label{fig:res_periodogram}
\end{figure}

\begin{figure}
    \centering
    \includegraphics[width=\columnwidth]{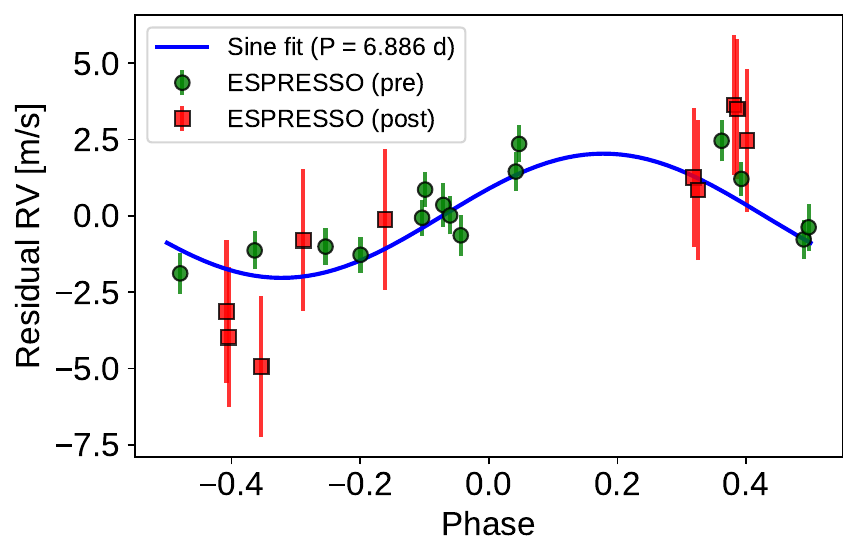}
    \caption{RV residuals phase-folded at the period corresponding to the strongest peak in the residual periodogram. The blue curve shows the best-fitting sinusoidal model obtained from a weighted linear fit. Green circles and red squares correspond to ESPRESSO observations obtained before and after the instrumental upgrade, respectively. Error bars include the quadrature sum of the measurement uncertainties and the adopted instrumental jitter.}

    \label{fig:res_folded}
\end{figure}

To investigate the presence of additional periodic variability in the data, we computed a generalized Lomb--Scargle periodogram of the RV residuals obtained after subtracting the median posterior predictive deterministic component of the model with the highest marginal likelihood (WN+SE+QP kernel without drift terms; Sect.~\ref{sect.rvanalysis}). The resulting periodogram 
(Fig.~\ref{fig:res_periodogram}) reveals a peak at $P = 6.89 \pm 0.15$ days above the $0.1$ $\%$ FAP. This signal is not related to the stellar rotation period and no clear feature is seen at this period in the ESPRESSO window function (Fig.~\ref{fig:wf}). However, it could be related to the second harmonic of the alias produced by the orbital period of LHS~3844\,b and the sidereal-day sampling, which would be expected at $P_{\rm alias} = 6.47 \pm 0.30$ days. This value lies within $\sim1.5\sigma$ of the detected peak. We also note that the same peak is present in the residuals of all other models considered in Sect.~\ref{sect.rvanalysis}. The RV residuals after subtracting the dominant periodic component, folded at this period, are shown in Fig.~\ref{fig:res_folded}. The periodogram also displays a secondary peak at $P = 14.85 \pm 0.72$ days, which corresponds to the first harmonic of the main signal and drops below the FAP thresholds once the 6.89-day periodic component is removed from the data.

Motivated by this result, we tested an extended model including an additional periodic signal. We adopted the same Bayesian framework and modelling choices described in Sect.~\ref{sect.rvanalysis}, starting from the configuration with highest marginal likelihood (WN+SE+QP $-$ $d_0$) and adding a second Keplerian component with circular orbit. The orbital phase parameter was assigned a uniform prior, while the period was given a moderately informative prior, modeled as a normal distribution in $\ln P$ centered on the period indicated by the residual periodogram ($\sim6.886$ days) with a dispersion of $\sigma=1$ in log-period space.

\begin{figure*}
    \centering
    \includegraphics[width=\textwidth]{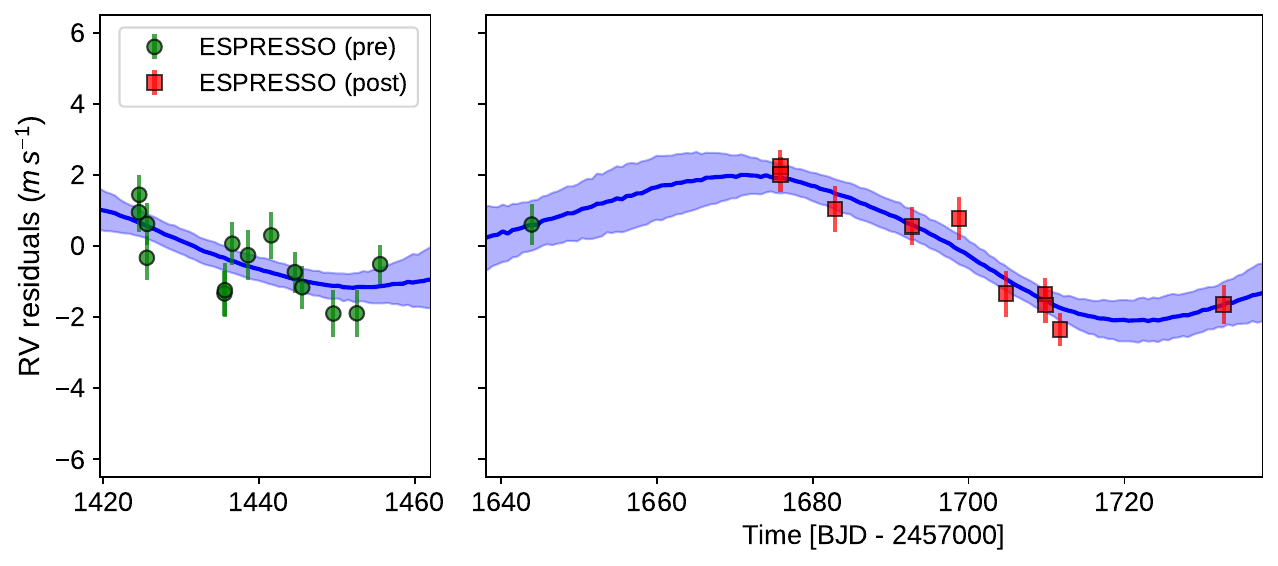}
    \caption{RV residuals of \object{LHS~3844} after subtracting the per-instrument offsets and the two orbital components, sampled from the posterior distribution of the model described in Sect. \ref{sect:secondsignal}. The blue curve represents the median of the Gaussian Process (GP) predictive distribution, computed over the full posterior of the GP hyperparameters and marginalized over the linear parameters and orbital period. The shaded region denotes the 68\% confidence interval. Data points correspond to ESPRESSO observations taken before (green circles) and after (red squares) the instrumental upgrade. Error bars include both the reported measurement uncertainties and the empirically derived per-instrument jitter, added in quadrature.}    
    \label{fig:GP_fit_2p}
\end{figure*}

The Markov chains converge well ($\hat{R} < 1.01$ for all parameters), indicating satisfactory mixing and convergence. The model recovers a second periodic signal with $P_2 = 6.694 \pm 0.007$ days and semi-amplitude $K_2 = 3.0 \pm 0.4$ m\,s$^{-1}$, corresponding to a minimum mass of $M_{2}\sin i = 2.5 \pm 0.4$ $M_\oplus$. In this configuration, the inferred semi-amplitude of LHS~3844\,b increases to $K = 7.85 \pm 0.49$ m\,s$^{-1}$. This would imply a larger planetary mass and would place LHS~3844\,b more firmly along the Earth-like composition curve in the mass–radius diagram of Fig.~\ref{fig:MR}. In contrast to the models discussed in Sect.~\ref{sect.rvanalysis}, the noise budget becomes strongly dominated by the periodic component of the Gaussian Process. The amplitudes of the jitter terms and of the white-noise and squared-exponential components are all consistent with zero within uncertainties, while the quasi-periodic component remains significantly non-zero at the level of $\sim$1–4 ms$^{-1}$. Its periodicity hyperparameter converges to $\mathcal{P}_{\rm QP} = 133^{+14}_{-19}$ days, consistent with the prior expectation based on the stellar rotation period. The corresponding GP predictive model is shown in Fig.~\ref{fig:GP_fit_2p}, where the correlated variability at the stellar rotation timescale becomes apparent once both periodic components are removed from the data.

The persistence of the $\sim6.9$-day signal across the residuals of all tested models, together with its absence in the ESPRESSO window function and in the periodograms of the stellar activity indicators (Fig.~\ref{fig:GLSrv}), raises the possibility that it may correspond to an additional planetary companion in the system. If interpreted as such, the signal would be consistent with a non-transiting planet LHS~3844\,c on a short-period orbit. This scenario is not implausible in the context of ultra-short-period (USP) systems, whose architectures often show enhanced mutual inclinations relative to longer-period compact systems, potentially leading to non-transiting companions \citep{dai2018}. 

However, the statistical support for this additional signal remains inconclusive. The marginal likelihood of the two-planet model is slightly higher than that of the single-planet models, but this comparison is affected by the informative prior adopted for the period of the second signal. Because this prior is centered on the detected peak, it likely increases the resulting model evidence. 
In addition, the detected period lies within $\sim1.5\sigma$ of the value expected for the second harmonic of an alias involving the orbital period of LHS~3844\,b and the sidereal-day sampling, which further complicates its interpretation. For this reason, we do not include this model in the evidence-weighted averaging used to derive our final estimates of $K$ and the other planetary parameters.\footnote{We also experimented with a log-uniform prior on the period of the second signal between 1 and 20 days. In this case the Markov chains diverged rapidly, and we therefore adopted the more restrictive prior described above.} In addition, the current dataset comprises only 25 radial-velocity measurements, limiting our ability to robustly disentangle additional periodic signals from stellar variability and correlated noise. Additional RV observations will therefore be required to confirm whether the $\sim6.9$-day signal corresponds to a second planet in the LHS~3844 system or arises from residual stellar or instrumental effects.

\subsection{Interior characterisation}
\label{sect.interior}

We quantify the confidence regions of possible interior properties using the approach in \citep{dorn_can_2015,dorn_generalized_2017} with recent updates from \citep{dorn_hidden_2021, luo_majority_2024}. We use a surrogate-accelerated Bayesian inference framework \citep{DeWringer2025}, which replaces the computationally expensive physics-based forward model with a fast polynomial chaos-Kriging (PCK) surrogate directly within a Markov chain Monte Carlo (MCMC) sampling loop \citep{schobi2015polynomial,marelli2014uqlab}. For our inference, the surrogate model provides high quality fits with R-squared values (coefficient of determination) of 0.98 and 0.99 for planetary mass and radius, respectively. Also, root mean square errors are well below observational uncertainties 0.03 and 0.004 for planetary mass and radius, respectively. Those errors of the model uncertainty are accounted for in the likelihood function.

The interior model assumes a rocky interior with the addition of water that is distributed between the different parts of a planet. The model consists of three layers: an iron-dominated core, a silicate mantle, and a H$_2$O steam atmosphere. We assume an adiabatic temperature profile for the core and mantle with possible temperature jumps at the core-mantle-boundary depending on melt temperatures. Further, we allow for both liquid and solid phases in those two layers.
For liquid iron and iron alloys we use the equation of state (EOS) by \citet{luo_majority_2024}. For solid iron, we use the EOS for hexagonal close packed iron \citep{hakim_new_2018,miozzi_new_2020}. 
For pressures below $\approx 125\,$GPa, the solid mantle mineralogy is modelled using the thermodynamical model \textsc{Perple\_X} \citep{connolly_geodynamic_2009} considering the system of MgO, SiO$_{2}$, and FeO. At higher pressures we define the stable minerals \textit{a priori} and use their respective EOS from various sources \citep{hemley_constraints_1992,fischer_equation_2011,faik_equation_2018,musella_physical_2019}.
The liquid mantle is modelled as a mixture of Mg$_2$SiO$_4$, SiO$_2$ and FeO \citep{melosh_hydrocode_2007,faik_equation_2018,ichikawa_ab_2020,stewart_shock_2020}, and mixed using the additive volume law. 

Water can be added to the mantle and core melts, depending on its solubility and partitioning behaviour, for which we follow \citep{dorn_hidden_2021, luo_majority_2024}. The addition of water reduces the density, for which we follow \citet{bajgain_structure_2015} and decrease the melt density per wt\% water by $0.036$ g cm$^{-3}$. For small water mass fractions, this reduction is nearly independent of pressure and temperature. The addition of water in core melts lowers the density as described in \citep{luo_majority_2024}. The effect of dissolved water on melting temperature is accounted for.
Beyond dissolved water in the deep interior, water can be in solid, supercritical and gas phase, for which we employ the EOS compilation in \citet{haldemann_aqua_2020}. 

The transit radius of a planet is assumed to be at a pressure of $P_{\rm Transit}=1$ mbar. This is a simplification as the transit radius depends on temperature, however, the effect on the planet of interest is small. The thermal profile is assumed to be fully adiabatic, except for pressures less than the pressure at the tropopause (here fixed at 0.1 bar) where we keep an isothermal profile that equals the equilibrium temperature. We use an equilibrium temperature of $737 \pm 25$K, assuming an albedo of 0.3.

\begin{figure}
    \centering
    \includegraphics[width=.8\columnwidth]{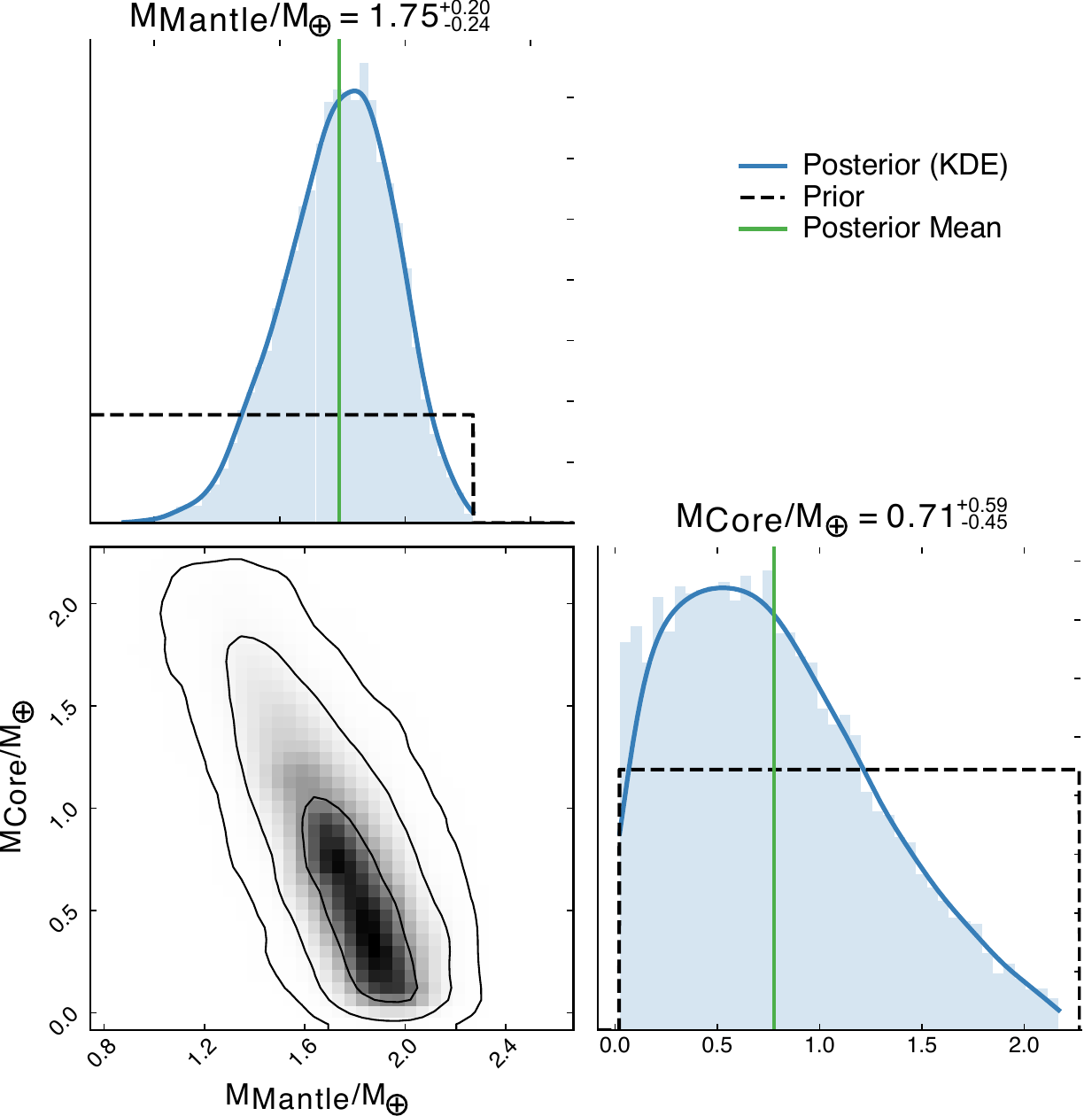}
    \caption{Marginalized posterior distribution of interior parameters in 1-D and 2-D. The shown posterior is calculated assuming a purely rocky interior with a fixed iron-free mantle composition and is constrained by planet mass, radius and equilibrium temperature.}
    \label{fig:interior_M-R_rocky}
\end{figure}

\begin{table}
\centering
\caption{Prior parameter distribution for the interior characterisation, assuming a purely rocky composition. $\mathcal{U}(a, b)$ is a uniform distribution between
$a$ and $b$.}\label{tab:comppriors1}
\renewcommand{\arraystretch}{1.2}

\begin{tabular}{lcc}
    \hline\hline
     Parameter & Prior dsitribution \\

    \hline \\[-6pt]%
    $M_\mathrm{mantle}$ & $\mathcal{U}(0.01\;M_{p}, 0.99\; M_{p}$)   \\
    $M_\mathrm{core}$  & $\mathcal{U}(0.01\;M_{p}, 0.99\; M_{p}$)   \\
    \bottomrule
\end{tabular}

\end{table}

We first consider a purely rocky interior, where we fix an iron-free mantle and use a simple uniform prior (Table \ref{tab:comppriors1}). Figure \ref{fig:interior_M-R_rocky} shows the 1D and 2D posterior distributions. In that case, the core mass fraction is well constrained to 0.29, which is similar to Earth's 0.325 estimate. There is a high posterior probability for core mass fractions lower than the Earth's value (see Figure \ref{fig:interior_M-R_rocky}), which is reflected in the lower mean density compared to an Earth-like composition. The possibility of an iron-rich super-Mercury interior is not supported by our new mass estimates \citep{kane2020}. A purely rocky interior, as assumed here, with the absence of any thick atmosphere is disfavoured given atmospheric observations \citep{diamondlowe2020}.

We also tested for a more general interior composition that allows for water distributed between core, mantle, and surface as well as a variable mantle composition (see Appendix \ref{fig:interior_M-R}). For this, we list the prior in Table \ref{tab:comppriors} in the Appendix. This more general composition comes with more degeneracy and we add the constraint that the bulk Fe/Si ratio is similar to Earth with $1.69 \pm 0.35$. We find that possible global water masses reach few percents ($0.01^{+0.02}_{-0.01}\%$), although the interior is also consistent in this case with a dry interior. The core mass fraction has a slightly higher estimate of 0.3 in that case, compensating for the addition of water.

In summary, the mass-radius data support a purely rocky interior with a core mass fraction that is similar to or tendentially smaller than that of Earth. Mass-radius data alone cannot distinguish existence or absence of a volatile layer on top of the rocky interior; however, such a volatile layer is disfavoured by phase curve measurements \citep{kreidberg2019}, transmission spectroscopy \citep{diamondlowe2020}, and atmospheric erosion considerations \citep{diamond-lowe2021}.

\section{Conclusion \label{sect.conclusion}}

We have presented the first dynamical mass determination for the ultra-short-period planet LHS~3844\,b, based on 25 high-resolution ESPRESSO spectra. Because the scope of the paper is to determine the mass of the planet we focused our efforts on the RV data. The information of the transits is included as priors on model parameters. 

To ensure a robust inference we adopted a fully Bayesian modelling framework and explored 15 competing radial-velocity models combining different Gaussian Process covariance kernels and polynomial drift prescriptions. Marginal likelihoods were computed for all models and used for Bayesian model averaging and evidence-weighted parameter estimation. As a consistency check, we also derived analytical solutions under the Maximum Likelihood Type-II approach, obtaining compatible mass estimates. 

The planetary signal is robustly detected throughout all models. From our marginalized, evidence-weighted posterior samples we derive a minimum mass of $M_p\sin i = 2.27 \pm 0.23\ M_\oplus$
and, combining with the transit inclination and the adopted stellar mass, an absolute mass consistent with this minimum mass. The resulting bulk density, $\rho = 5.67 \pm 0.65\ \mathrm{g\,cm^{-3}}$, is consistent with a predominantly rocky composition.

Model comparison (marginal likelihoods) systematically favours kernels that include periodic or quasi-periodic components and disfavours models with additional long-term linear/quadratic drifts. Models without drift ($d_0$) dominate the posterior model probability across kernel choices, and three kernel configurations (\texttt{WN+QP}, \texttt{WN+SE+QP}, \texttt{WN+SE+Per}) contribute the bulk of the evidence. Importantly, the inferred \(K\) is stable across every model, demonstrating that the mass estimate is not sensitive to the precise choice of kernel or drift model.

The analysis of the RV residuals also reveals a periodic signal at $\sim6.9$ days that appears consistently across the residuals of all tested models. An extended two-planet model recovers this signal with a semi-amplitude of $\sim3$ m\,s$^{-1}$, but its statistical support remains inconclusive given the limited number of observations and the possible impact of sampling and prior assumptions on the inference. While this signal may hint at the presence of an additional non-transiting companion in the system, further RV measurements will be necessary to confirm its planetary nature.

Interior-structure inference indicates a predominantly rocky composition for LHS 3844\,b with a core mass fraction similar to or slightly lower than Earth. Allowing for water-bearing interiors yields only trace water mass fractions within the posterior; a thick volatile envelope is disfavoured by the available constraints and by previous atmospheric observations \citep[e.g.][]{kreidberg2019,diamondlowe2020}.

Taken together, the results robustly establish LHS~3844\,b as a terrestrial, ultra-short-period planet with a well-constrained mass and density. Although the current RV dataset suffices to measure the planetary mass at the $\sim$10\% level, further improvements in precision and/or cadence (and additional epochs) would help to reduce the remaining uncertainty on $K$, better characterise the GP hyperparameters associated with stellar variability, and test for lower-amplitude companions. 
Finally, the precise mass and density reported here strengthen the scientific case for continued atmospheric and surface characterisation of LHS~3844\,b with JWST. Our dynamical constraints provide a necessary anchor for interpreting emission and phase-curve observations and for placing the planet in the broader context of terrestrial exoplanet composition and evolution.

During the preparation of this manuscript, we became aware of an independent and contemporaneous analysis by Nagel et al. (University of Göttingen) also reporting a mass measurement of LHS~3844~b using ESPRESSO and CRIRES+ data.

\begin{acknowledgements}

N.A-D. acknowledges the support of FONDECYT project 1240916. 

C.D acknowledges support from the Swiss National Science Foundation under grant TMSGI2\_211313. This work has been carried out within the framework of the NCCR PlanetS supported by the Swiss National Science Foundation under grant 51NF40\_205606. 

XB, XD \& TF acknowledge funding from the French ANR under contract number ANR18-CE310019 (SPlaSH), and the French National Research Agency in the framework ofWeighing the mass of LHS 3844 b the Investissements d’Avenir program (ANR-15-IDEX-02), through the funding of the “Origin of Life” project of the Grenoble-Alpes University.
\end{acknowledgements}

\bibliographystyle{aa}
\bibliography{LHS3844_biblio}

\begin{appendix}

\section{Orbital model \label{appendix_orbital_model}}

The effect of the planet's gravitational influence is modeled using a single Keplerian function, assuming one orbiting companion. The Keplerian signal is parameterised by the orbital period $P$, the semi-amplitude $K$, the eccentricity $e$, the argument of periastron $\omega$, and the mean anomaly $M_0$. The RV contribution from the planet at time $t_i$ is:

\begin{equation}
    v_\mathrm{kep} = K \left[\cos\left(f_i + \omega\right) + e \cos(\omega)\right]   
\label{eq:RV}
\end{equation}

\noindent where $f_i = f(t_i; P, e, \omega, M_0)$ is the true anomaly at time $t_i$. 

\subsection{Circular orbit \label{appendix_orbital_model_circular}}
Assuming a circular orbit, which corresponds to $e = 0$, the dependence on the argument of periastron $\omega$ is eliminated, and $f_i$ becomes equal to $M_0$. Then, the radial velocity contribution simplifies to:
\[
v_\mathrm{kep} = K\cos(\omega) \cos\left(M_0\right) + K\sin(\omega)\sin\left(M_0\right),
\]

Since we can write the mean anomaly as $M_0 = \frac{2\pi}{P}(t-\tau)$, with $\tau$ being the periastron passage time, we can rewrite the contribution from the circular orbit as

\[
v_\mathrm{kep} = \beta_0 \sin \left(\frac{2\pi t}{P} \right) + \beta_1 \cos \left(\frac{2\pi t}{P} \right),
\]

\noindent where $\beta_0 = -K\sin\left(\omega - \frac{2\pi}{P}\tau\right)$ and $\beta_1 = K\cos\left(\omega- \frac{2\pi}{P}\tau\right)$. This way we can recover the amplitude $K$ of the RV from the coefficients of the
model, through $K = \sqrt{\beta_0^2 + \beta_1^2}$.

\subsection{Mildly eccentric orbit \label{appendix_orbital_model_ecc}}

We can lift the assumption of circular orbit while keeping the linear nature of the model and all of its advantages by expanding the Keplerian expression of the radial velocity (Eq.~\ref{eq:RV}) as a series of the orbital eccentricity. To first order, we obtain \citep{lucy1971}:

\begin{equation}
    v_\mathrm{kep} = K\cos(M_0) + eK\cos(2M_0 - \omega) + O(e^2).
\end{equation}

\noindent Expanding the second term and neglecting terms of order $O(e^2)$ or higher, this simplifies to:

\begin{equation}
    V = V_0 + K'\cos(M_0) + eK'\cos(\omega)\sin(2M_0) + eK'\sin(\omega)\cos(2M_0),
\end{equation}

\noindent where $K'$ is related to $K$ by

\begin{equation}
    K' = \frac{\cos\left(\omega-\frac{2\pi}{P}\tau\right)}{\cos\left( \frac{2\pi}{P}\tau\right)}K.
\end{equation} 

\par Since $2M_0(P) = M_0(P/2)$, the eccentricity-dependent terms introduce periodic components at the first harmonic $P/2$ instead of the orbital period $P$. Thus, by fitting these periodic terms to the data, one can estimate their coefficients and infer the orbital eccentricity.

\section{Maximum likelihood type-II robustness analysis}
\label{appendix:ml2}

As an additional robustness check of the planetary signal characterized in Sect.~\ref{sect.rvanalysis}, we implemented an alternative inference scheme based on the Maximum Likelihood Type-II approach \citep[ML-II;][]{rasmussen}. In this framework, the likelihood marginalized over the linear parameters is maximized with respect to the hyperparameters, which are fixed at the values that maximize the marginal likelihood of the model. 

We adopted the same deterministic and noise models described in Sect.~\ref{sect.rvanalysis}, assuming Gaussian priors for the linear parameters. However, in contrast to the full Bayesian MCMC analysis, the planetary period $P$ and transit epoch $t_0$ were fixed to their transit-derived values. Under these assumptions, and given that the measurement errors are modeled as multivariate normal, the marginalization over the linear parameters can be performed analytically.

For each of the five Gaussian Process kernels considered in the main analysis, we applied an ML-II optimisation procedure. First, the marginal likelihood was maximized with respect to the GP hyperparameters of the selected kernel. Once the optimal hyperparameters were obtained, the model was conditioned on these values and a Bayesian linear regression was performed to derive the posterior distribution of the linear parameters, including the planetary semi-amplitude $K$ and instrumental offsets. The semi-amplitude \( K \) values and the corresponding log-marginal-likelihoods for each kernel model are reported in Table~\ref{tab:MLII_K_comparison}.

\begin{figure*}
    \sidecaption
    \includegraphics[width=12cm]{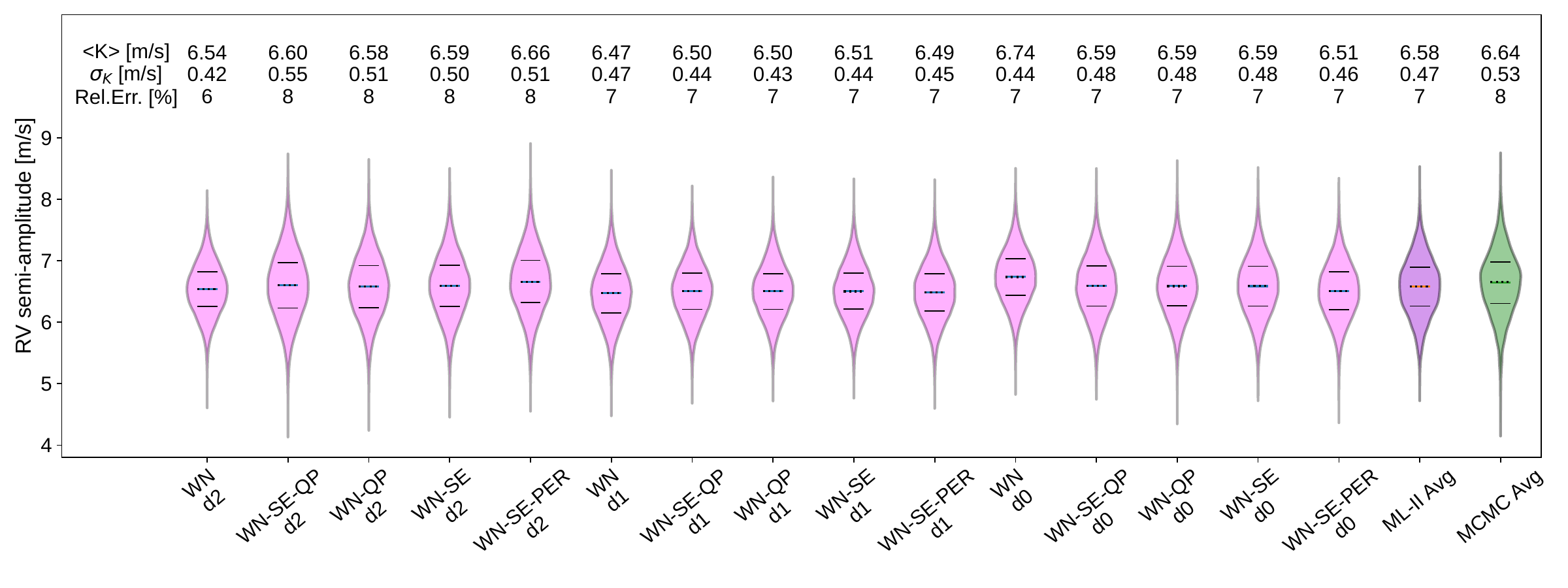}
    \caption{Comparison of the inferred RV semi-amplitude ($K$) across the ML-II analytical models, ordered by increasing marginal likelihood. For each model, the violin represents a sampling distribution obtained by drawing from a normal distribution with mean $K$ and standard deviation $\sigma_K$ derived from the analytical ML-II results. The mean (solid line), median (dotted line), and 1st/3rd quartiles are indicated. The second violin on the right corresponds to the ML-II model-averaged estimate, constructed by weighting each model according to its marginal likelihood, while the rightmost violin shows the posterior model-averaged estimate derived from the MCMC analysis. Numerical annotations report the mean, standard deviation, and relative uncertainty of $K$ for each case.}
    \label{fig:semiamplitude_mlII}
\end{figure*}

To facilitate comparison with the fully Bayesian results, sampling distributions for $K$ were generated for each ML-II model by drawing from normal distributions centered on the analytical estimates and with standard deviations given by their corresponding uncertainties. The models were ordered according to increasing marginal likelihood, and a model-averaged estimate was constructed by weighting each model by its marginal likelihood. The resulting distributions are shown in Fig.~\ref{fig:semiamplitude_mlII}. The inferred values of $K$ are fully consistent with those obtained from the MCMC analysis presented in the main text, confirming the robustness of the detected planetary signal under this alternative inference framework.

\begin{table*}[ht]
\centering
\caption{Results obtained from the ML type-II models described in Sect.~\ref{appendix:ml2}. 
}
\begin{tabular}{|l|c|c|c|c|c|}
\hline
\textbf{Parameter / Metric} & \textbf{White} & \textbf{SE-White} & \textbf{QP-White} & \textbf{SE-QP-White} & \textbf{SE-Per-White} \\
\hline

\multicolumn{6}{|c|}{\textbf{Drift $d_0$}} \\
\hline
RV semi-amplitude $K$ $(ms^{-1})$ 
& $6.74 \pm 0.44$ 
& $6.59 \pm 0.48$ 
& $6.59 \pm 0.48$ 
& $6.59 \pm 0.48$ 
& $6.51 \pm 0.46$ \\

Marginal likelihood $\ln \mathcal{L}$ 
& $-58.93$ 
& $-58.30$ 
& $-58.30$ 
& $-58.30$ 
& $-57.99$ \\
\hline

\multicolumn{6}{|c|}{\textbf{Drift $d_1$}} \\
\hline
RV semi-amplitude $K$ $(ms^{-1})$ 
& $6.47 \pm 0.47$ 
& $6.50 \pm 0.44$ 
& $6.50 \pm 0.43$ 
& $6.51 \pm 0.44$ 
& $6.49 \pm 0.45$ \\

Marginal likelihood $\ln \mathcal{L}$ 
& $-65.27$ 
& $-63.09$ 
& $-63.10$ 
& $-63.12$ 
& $-62.92$ \\
\hline

\multicolumn{6}{|c|}{\textbf{Drift $d_2$}} \\
\hline
RV semi-amplitude $K$ $(ms^{-1})$ 
& $6.54 \pm 0.42$ 
& $6.59 \pm 0.50$ 
& $6.58 \pm 0.50$ 
& $6.60 \pm 0.55$ 
& $6.66 \pm 0.51$ \\

Marginal likelihood $\ln \mathcal{L}$ 
& $-76.17$ 
& $-74.69$ 
& $-74.72$ 
& $-74.78$ 
& $-74.46$ \\
\hline

\end{tabular}\label{tab:MLII_K_comparison}
{\captionsetup{font=small}
    \caption*{\textbf{Notes.} For each kernel, the table reports the radial velocity semi-amplitude $K$ as the mean $\pm$ standard deviation, and the corresponding marginal log-likelihood ($\ln \mathcal{L}$) of the model.}}
\end{table*}

\section{Additional plots and tables \label{appendix_plots_tables}}

\begin{figure*}
    \sidecaption
    \includegraphics[width=12cm]{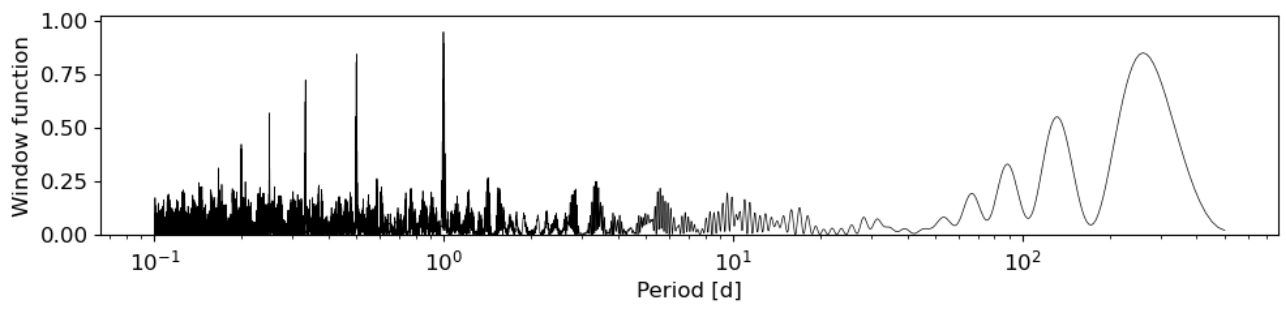}
    \caption{Window function of the ESPRESSO data, dominated by a strong peak at the sidereal day.}
    \label{fig:wf}
\end{figure*}

The ESPRESSO data described in Sect.~\ref{sect.data} are listed in Table~\ref{tab:data_summary}, and their window function is shown in Fig.~\ref{fig:wf}.

Further details on the radial-velocity modelling are provided in Tables~\ref{tab:nonlinear_parameters_kernel} and \ref{tab:linear_parameters_drift}, which summarize the posterior distributions of the non-linear and linear model parameters, respectively. 

\begin{figure*}
    \centering
    \includegraphics[width=0.9\textwidth]{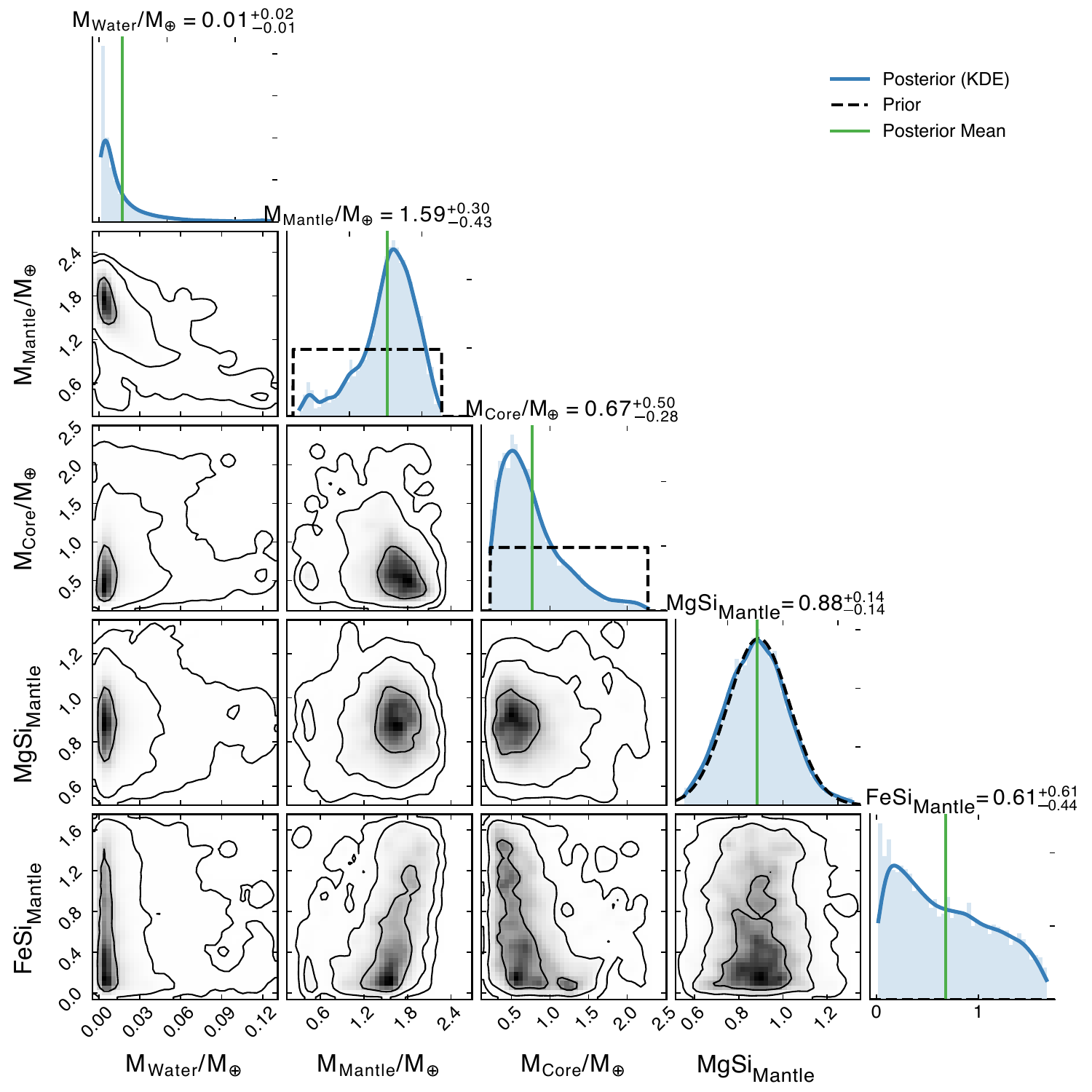}
    \caption{Marginalized posterior distribution of interior parameters in 1-D and 2-D. The shown posterior is calculated assuming a rocky interior with the possibility of water distributed between core, mantle, and surface. The mantle composition can vary in terms of Fe/Si and Mg/Si ratios. The posterior is constrained by planet mass, radius and equilibrium temperature, and an Fe/Si bulk ratios similar to Earth (see main text).}
    \label{fig:interior_M-R}
\end{figure*}

Finally, Table~\ref{tab:comppriors} lists the prior parameter distributions adopted in the interior modelling (described in Sect.~\ref{sect.interior}), while Fig.~\ref{fig:interior_M-R} shows the resulting marginalized posterior distributions.

\begin{table}
\centering
\caption{Prior parameter distribution for the interior characterisation. }
\label{tab:comppriors}
\renewcommand{\arraystretch}{1.2}

\begin{tabular}{lcc}
    \hline\hline
     Parameter & Prior distribution \\
    \hline \\[-6pt]%
    $f_\mathrm{water}$ & $\ln\mathcal{U}(10^{-3}\;M_{p}, 0.05\;M_{p})$ \\
    $M_\mathrm{mantle}$& $\mathcal{U}(0.1\;M_{p}, 0.9\; M_{p})$ \\
    $M_\mathrm{core}$  & $\mathcal{U}(0.1\;M_{p}, 0.9\; M_{p})$ \\
    Mg/Si$_\mathrm{mantle}$ & $\mathcal{N}(0.886, 0.143)$  \\
    Fe/Si$_\mathrm{mantle}$ & $\mathcal{U}(0.0, 1.69)$ \\
    \bottomrule
\end{tabular}

{\captionsetup{font=small}
    \caption*{\textbf{Notes.} $\mathcal{N}(\mu, \sigma)$ denotes a normal distribution with mean $\mu$ and standard deviation $\sigma$;  $\mathcal{U}(a,b)$ and $\ln\mathcal{U}(a,b)$ is a uniform, and log-uniform distribution, respectively, between $a$ and $b$.}}
\end{table}

\begin{table*}[ht]
\renewcommand{\arraystretch}{1.2}
\centering
\caption{Spectroscopic measurements used in this work.}
\label{tab:data_summary}
\resizebox{\textwidth}{!}{
\begin{tabular}{cccccccccc}
\hline
\textbf{Time (BJD-2450000)} & \textbf{Instrument} & \textbf{RV [km/s]} & \textbf{FWHM [km/s]} & \textbf{H$\alpha$} & \textbf{H$\beta$} & \textbf{H$\gamma$} & \textbf{Na D} & \textbf{S-index} \\
\hline
8424.6006 & ESPRESSO & $-10.5235 \pm 0.0006$ & $5.4565 \pm 0.0011$ & $0.08710 \pm 0.00017$ & $0.08585 \pm 0.00073$ & $0.1794 \pm 0.0028$ & $0.02571 \pm 0.00020$ & $2.278 \pm 0.067$ \\
8424.6330 & ESPRESSO & $-10.5205 \pm 0.0006$ & $5.4504 \pm 0.0011$ & $0.08665 \pm 0.00016$ & $0.08794 \pm 0.00071$ & $0.1956 \pm 0.0028$ & $0.02865 \pm 0.00021$ & $2.358 \pm 0.063$ \\
8425.6032 & ESPRESSO & $-10.5187 \pm 0.0006$ & $5.4518 \pm 0.0013$ & $0.09182 \pm 0.00019$ & $0.10508 \pm 0.00091$ & $0.2329 \pm 0.0037$ & $0.02208 \pm 0.00021$ & $2.561 \pm 0.084$ \\
8425.6392 & ESPRESSO & $-10.5186 \pm 0.0006$ & $5.4501 \pm 0.0012$ & $0.09157 \pm 0.00018$ & $0.10053 \pm 0.00083$ & $0.2101 \pm 0.0033$ & $0.02089 \pm 0.00019$ & $2.557 \pm 0.073$ \\
8435.5697 & ESPRESSO & $-10.5341 \pm 0.0006$ & $5.4550 \pm 0.0013$ & $0.08574 \pm 0.00019$ & $0.08478 \pm 0.00083$ & $0.1854 \pm 0.0032$ & $0.03004 \pm 0.00025$ & $2.501 \pm 0.099$ \\
8435.6228 & ESPRESSO & $-10.5313 \pm 0.0008$ & $5.4706 \pm 0.0018$ & $0.08885 \pm 0.00024$ & $0.09225 \pm 0.00111$ & $0.1936 \pm 0.0043$ & $0.05130 \pm 0.00041$ & $3.307 \pm 0.210$ \\
8436.5798 & ESPRESSO & $-10.5295 \pm 0.0006$ & $5.4458 \pm 0.0013$ & $0.08801 \pm 0.00019$ & $0.09442 \pm 0.00084$ & $0.1914 \pm 0.0031$ & $0.02480 \pm 0.00021$ & $2.138 \pm 0.061$ \\
8438.5988 & ESPRESSO & $-10.5205 \pm 0.0007$ & $5.4525 \pm 0.0016$ & $0.08679 \pm 0.00022$ & $0.08488 \pm 0.00094$ & $0.1762 \pm 0.0036$ & $0.03256 \pm 0.00029$ & $2.485 \pm 0.116$ \\
8441.5769 & ESPRESSO & $-10.5310 \pm 0.0007$ & $5.4600 \pm 0.0015$ & $0.08688 \pm 0.00021$ & $0.08075 \pm 0.00084$ & $0.1529 \pm 0.0028$ & $0.02773 \pm 0.00025$ & $1.479 \pm 0.062$ \\
8444.5989 & ESPRESSO & $-10.5216 \pm 0.0006$ & $5.4579 \pm 0.0012$ & $0.09756 \pm 0.00019$ & $0.10889 \pm 0.00082$ & $0.1821 \pm 0.0024$ & $0.02464 \pm 0.00020$ & $1.673 \pm 0.038$ \\
8445.5572 & ESPRESSO & $-10.5215 \pm 0.0006$ & $5.4503 \pm 0.0013$ & $0.08778 \pm 0.00019$ & $0.07707 \pm 0.00071$ & $0.1360 \pm 0.0021$ & $0.02378 \pm 0.00021$ & $1.229 \pm 0.037$ \\
8449.5542 & ESPRESSO & $-10.5292 \pm 0.0007$ & $5.4612 \pm 0.0014$ & $0.09067 \pm 0.00020$ & $0.09386 \pm 0.00092$ & $0.2080 \pm 0.0037$ & $0.02179 \pm 0.00022$ & $2.643 \pm 0.109$ \\
8452.5566 & ESPRESSO & $-10.5264 \pm 0.0007$ & $5.4479 \pm 0.0015$ & $0.08846 \pm 0.00021$ & $0.08696 \pm 0.00084$ & $0.1598 \pm 0.0025$ & $0.02529 \pm 0.00023$ & $1.539 \pm 0.029$ \\
8455.5597 & ESPRESSO & $-10.5272 \pm 0.0005$ & $5.4444 \pm 0.0011$ & $0.09501 \pm 0.00017$ & $0.10555 \pm 0.00079$ & $0.2090 \pm 0.0029$ & $0.01887 \pm 0.00016$ & $2.824 \pm 0.063$ \\
8643.9056 & ESPRESSO & $-10.5339 \pm 0.0006$ & $5.4189 \pm 0.0012$ & $0.08872 \pm 0.00018$ & $0.08947 \pm 0.00079$ & $0.2073 \pm 0.0030$ & $0.01025 \pm 0.00013$ & $1.956 \pm 0.070$ \\
8675.8201 & ESPRESSO+ & $-10.5258 \pm 0.0005$ & $5.4498 \pm 0.0009$ & $0.08167 \pm 0.00014$ & $0.07488 \pm 0.00059$ & $0.1568 \pm 0.0020$ & $0.01024 \pm 0.00011$ & $1.833 \pm 0.048$ \\
8675.8539 & ESPRESSO+ & $-10.5251 \pm 0.0005$ & $5.4431 \pm 0.0009$ & $0.08076 \pm 0.00014$ & $0.07251 \pm 0.00059$ & $0.1584 \pm 0.0021$ & $0.00972 \pm 0.00011$ & $1.878 \pm 0.052$ \\
8682.8435 & ESPRESSO+ & $-10.5232 \pm 0.0006$ & $5.4544 \pm 0.0016$ & $0.08183 \pm 0.00020$ & $0.06227 \pm 0.00070$ & $0.1113 \pm 0.0018$ & $0.01551 \pm 0.00019$ & $0.825 \pm 0.039$ \\
8692.7410 & ESPRESSO+ & $-10.5167 \pm 0.0005$ & $5.4484 \pm 0.0011$ & $0.08739 \pm 0.00017$ & $0.08942 \pm 0.00078$ & $0.1857 \pm 0.0027$ & $0.01899 \pm 0.00017$ & $2.041 \pm 0.068$ \\
8698.7535 & ESPRESSO+ & $-10.5173 \pm 0.0006$ & $5.4542 \pm 0.0013$ & $0.08217 \pm 0.00018$ & $0.07705 \pm 0.00081$ & $0.1774 \pm 0.0030$ & $0.02128 \pm 0.00020$ & $1.779 \pm 0.087$ \\
8704.8156 & ESPRESSO+ & $-10.5218 \pm 0.0006$ & $5.4326 \pm 0.0015$ & $0.08570 \pm 0.00020$ & $0.08488 \pm 0.00093$ & $0.1893 \pm 0.0035$ & $0.02269 \pm 0.00023$ & $2.094 \pm 0.109$ \\
8709.8159 & ESPRESSO+ & $-10.5153 \pm 0.0004$ & $5.4426 \pm 0.0009$ & $0.08799 \pm 0.00013$ & $0.08434 \pm 0.00055$ & $0.1564 \pm 0.0016$ & $0.01913 \pm 0.00013$ & $1.403 \pm 0.027$ \\
8709.8634 & ESPRESSO+ & $-10.5156 \pm 0.0005$ & $5.4575 \pm 0.0011$ & $0.08706 \pm 0.00016$ & $0.07451 \pm 0.00059$ & $0.1360 \pm 0.0016$ & $0.02572 \pm 0.00018$ & $1.199 \pm 0.031$ \\
8711.7255 & ESPRESSO+ & $-10.5208 \pm 0.0005$ & $5.4479 \pm 0.0010$ & $0.08502 \pm 0.00014$ & $0.07552 \pm 0.00055$ & $0.1332 \pm 0.0015$ & $0.01852 \pm 0.00014$ & $1.197 \pm 0.027$ \\
8732.7301 & ESPRESSO+ & $-10.5339 \pm 0.0006$ & $5.4475 \pm 0.0011$ & $0.09579 \pm 0.00017$ & $0.11206 \pm 0.00088$ & $0.2316 \pm 0.0031$ & $0.02289 \pm 0.00019$ & $2.763 \pm 0.076$ \\
\hline
\end{tabular}
}
{\captionsetup{font=small}
    \caption*{\textbf{Notes.} "ESPRESSO" and "ESPRESSO+" labels correspond to observations taken before and after the fibre link upgrade, respectively.}}
\end{table*}

\begin{table*}[ht]
\centering
\caption{Means and standard deviations of the posterior distributions for the non-linear parameters, marginalized over drift models with a common kernel via ancestral sampling, using the marginal likelihood as model weights.}

\resizebox{\textwidth}{!}{
\begin{tabular}{|l|c|c|c|c|c|}
\hline
\textbf{Parameter / Hyperparameter} & \textbf{White} & \textbf{SE-White} & \textbf{QP-White} & \textbf{SE-QP-White} & \textbf{SE-Per-White} \\ \hline

Jitter $ \ln \sigma_{Jt}$ (pre) ($m s^{-1}$) 
& -1.94 $\pm$ 2.38 
& -2.04 $\pm$ 2.36 
& -2.25 $\pm$ 2.36 
& -2.30 $\pm$ 2.38 
& -2.25 $\pm$ 2.36 \\ 

Jitter $\ln \sigma_{Jt}$ (post) ($m s^{-1}$) 
& 0.73 $\pm$ 1.36 
& 0.61 $\pm$ 1.53 
& -0.15 $\pm$ 2.19 
& -0.22 $\pm$ 2.22 
& -0.18 $\pm$ 2.18 \\ 

White-noise amplitude $\ln A_\mathrm{white}$ ($m s^{-1}$) 
& -1.26 $\pm$ 2.25 
& -1.15 $\pm$ 2.21 
& -0.94 $\pm$ 2.13 
& -0.88 $\pm$ 2.08 
& -0.87 $\pm$ 2.09 \\ 

Covariance amplitude $\ln A_\mathrm{SE}$ ($m s^{-1}$) 
& --- 
& -2.26 $\pm$ 2.68 
& --- 
& -2.38 $\pm$ 2.66 
& -2.28 $\pm$ 2.63 \\ 

Decay timescale $\ln \tau_\mathrm{SE}$ ($d$) 
& --- 
& 5.92 $\pm$ 0.83 
& --- 
& 5.94 $\pm$ 0.82 
& 5.94 $\pm$ 0.83 \\ 

Covariance amplitude $\ln A_\mathrm{QP/Per}$ ($m s^{-1}$) 
& --- 
& --- 
& -1.45 $\pm$ 2.64 
& -1.47 $\pm$ 2.63 
& -1.46 $\pm$ 2.63 \\ 

Smoothness parameter, $\ln \lambda_\mathrm{QP/Per}$ [-] 
& --- 
& --- 
& 0.04 $\pm$ 0.39 
& 0.04 $\pm$ 0.39 
& 0.03 $\pm$ 0.40 \\ 

Periodicity $\ln \mathcal{P}_\mathrm{QP/Per}$ ($d$) 
& --- 
& --- 
& 4.85 $\pm$ 0.19 
& 4.86 $\pm$ 0.19 
& 4.85 $\pm$ 0.19 \\ 

Decay timescale $\ln \tau_\mathrm{QP}$ ($d$) 
& --- 
& --- 
& 5.96 $\pm$ 0.80 
& 5.96 $\pm$ 0.79 
& --- \\   \hline
\end{tabular}
}
\label{tab:nonlinear_parameters_kernel}
{\captionsetup{font=small}
    \caption*{\textbf{Notes.} Parameters marked with an asterisk indicate the kernel amplitudes that dominate the covariance structure in each model (see discussion in Sect.~\ref{sect.linparam}).}}
\end{table*}

\begin{table*}[ht]
\centering
\caption{Means and standard deviations of the posterior distributions for the linear parameters, marginalized over kernel models with a common drift term via ancestral sampling, using the marginal likelihood as model weights.}
\begin{tabular}{|l|c|c|c|}
\hline
\textbf{Parameter} & \textbf{Drift: d0} & \textbf{Drift: d1} & \textbf{Drift: d2} \\ \hline
General offset $\gamma_0$ ($m s^{-1}$) & -1.12 $\pm$ 2.49 & -2.70 $\pm$ 3.14 & 0.42 $\pm$ 3.16 \\ 
ESPRESSO-post offset $\gamma_0'$ ($m s^{-1}$) & -0.25 $\pm$ 1.95 & 3.16 $\pm$ 3.37 & 6.67 $\pm$ 3.28 \\ 
Linear trend $\gamma_1$ ($m s^{-1} d^{-1}$) & --- & -0.02 $\pm$ 0.02 & -0.03 $\pm$ 0.02 \\ 
Quadratic trend $\gamma_2$ ($m s^{-1} d^{-2}$) & --- & --- & -0.0003 $\pm$ 0.0002 \\ 
RV semi amplitude $K$ ($m s^{-1}$) & 6.64 $\pm$ 0.53 & 6.41 $\pm$ 0.55 & 6.56 $\pm$ 0.53 \\ \hline
\end{tabular}
\label{tab:linear_parameters_drift}
\end{table*}

\end{appendix}

\end{document}